\documentclass[fleqn, preprint, number]{elsarticle} 
\usepackage[a4paper]{geometry}
\usepackage{graphicx} % Required for inserting images
\usepackage{longtable}
\usepackage{amsmath}
\usepackage{amssymb}
\usepackage{hyperref}
\usepackage[gen]{eurosym}
\usepackage{bbm} 
\usepackage{datetime2}
\usepackage{cases}
\usepackage[utf8]{inputenc}
\usepackage[T1]{fontenc}
\usepackage{nicefrac}
\usepackage{booktabs}
\begin{document}
%remove elsevier line
\makeatletter
\def\ps@pprintTitle{%
  \let\@oddhead\@empty
  \let\@evenhead\@empty
  \def\@oddfoot{\reset@font\hfil\thepage\hfil}
  \let\@evenfoot\@oddfoot
}
\makeatother

\title{Coordinated Trading Strategies for Battery Storage in Reserve and Spot Markets}  

\author[1]{Paul E. Seifert \corref{cor1}}
\ead{paules@ntnu.no}
\author[2]{Emil Kraft}
\author[1]{Steffen J. Bakker}
\author[1]{Stein-Erik Fleten}

\affiliation[1]{
            organization={Department of Industrial Economics and Technology Management, Norwegian University of Science and Technology},
            addressline={Alfred Getz' vei 3},
            city={7491 Trondheim},
            country={Norway}
            }
\affiliation[2]{
            organization={Institute for Industrial Production,
            Karlsruhe Institute of Technology},
            addressline={Hertzstr. 16}, 
            city={76187 Karlsruhe},
            country={Germany}}

\begin{abstract}
Battery storage systems play a crucial role in modern energy infrastructure by balancing fluctuations of renewable energy sources. However, price interdependencies between markets and time-coupling constraints of storage complicate daily bidding and operation across available electricity markets. Our research contributes to the literature on short-term electricity market operations by expanding it to frequency containment reserve, day-ahead, and intraday markets. Our contribution captures price and quantity uncertainty in a stochastic modelling approach that complies with the sequential price dependencies using a multidimensional Markov Chain. We use econometrics to segregate market sentiments from intrinsic stochasticity and analyse price drivers specific to each market to generate representative scenarios. The applicability of this method, even in exceptional market environments, is demonstrated in a case study of the German electricity market zone during the European energy crisis in 2022. We provide insights into the operation of battery storage and find that capacity reservation in the frequency containment reserve dominates day-ahead and intraday bidding during this time, indicating significant volume shifts between markets. Under cycling compensated market conditions, we observe that coordinated bidding across multiple markets can yield additional value and estimate it to be up to 12.5\,\% since battery storage can economically leverage price fluctuations.
\end{abstract}

\begin{highlights}
\item Explores coordinated bidding complexities and steps to build an SDDP-based trading strategy.
\item The battery storage in the case study operates in three markets (DA, ID and FCR).
\item At four-hour resolution, the FCR market dominates, with minimal coordination benefits.
\item Modified prices enable coordinated bidding and increase revenues by up to 12.5\,\%.
\end{highlights}

% Keywords
% Each keyword is separated by \sep

\maketitle
\section{Introduction}
Renewable energy sources (RES) supersede controllable power plants in the electricity system due to their economic competitiveness and the need to reduce carbon emissions \citep{ramsebner_sector_2021, kuik_competitive_2019}. However, RES rely on weather conditions and are often located far from demand centres. This intensifies inflexibility issues in space and time \citep{lund_review_2015, edenhofer_economics_2013}. Due to RES's limited predictability and semi-dispatchability, it is difficult to increase RES shares above current levels and reduce controllable generation without reducing system stability \citep{bublitz_survey_2019}. Therefore, it is expected that demands for balancing capacity and reserves increase to ensure the system balance. However, field studies show that market design improvements and the introduction of the Intraday (ID) market have led to the opposite effect of reduced balancing needs \citep{koch_short-term_2019}. The authors' analysis is limited to the ID market; however, there are more markets in which to participate. Synergies from market integration and sector coupling are well-studied \citep{fridgen_holistic_2020}. The recent cost reductions and the ability to store electricity with short lead times have made battery storage a promising technology for this cross-market integration and can help reduce the need for backup capacity in the future. However, integrating battery storage into multiple markets is challenging but necessary: older studies report revenue stacking of multiple markets necessary to operate battery storage profitable \citep{castagneto_gissey_market_2018, heredia_optimal_2018}. Given the recent cost reductions and expected build-out in battery storage, having a competitive edge is essential for market participants to operate battery storage profitably in the future.

% Market coordination
%Simultaneously, participation in different electricity markets provides multiple revenue streams. This coordination can yield higher revenue by leveraging synergies from opposing market movements. 
Market coordination refers to a process of coordinating decisions across multiple markets, taking into account the expectations of subsequent markets \citep{lohndorf_value_2022}. Instead of a series of individual optimisations, market coordination determines the best possible decisions across all markets while considering the uncertainty of future actions and parameters at the time of decision-making.
% Coordination of a battery storage
%In the context of battery storage, bidding for prices and quantities represents this uncertainty in decision-making. 
The scheduling of markets has inherent time gaps between bidding, market clearing, and resource deployment. Battery storage operators can take advantage of these gaps to benefit from price changes and diversify risks through portfolio optimization. However, the complexity of multi-stage decision-making with time constraints complicates the development of optimal trading strategies as calculations suffer heavily from the curse of dimensionality, often resulting in oversimplified approaches, such as assuming perfect foresight or simplified uncertainty.

Given these challenges and the successful usage of Stochastic Dual Dynamic Programming (SDDP) in managing the uncertainty of hydro assets, the main research question addressed in this paper is the following. Is the use of SDDP adequate to depict the storage operator's problem for coordinated bidding in three markets under given computational and technical limitations while following the German market schedule? This paper aims to calculate optimal strategies for battery storage operators in coordinated markets from known uncertainty and provide insights into overcoming the inherent complexities and uncertainties.

% Contribution
In this paper, we develop a scalable method for multi-market battery storage bidding under uncertainty, which considers the sequential timing structure of real markets and utilises SDDP to solve the bidding problem. We consider the intricate relationships between the times of bid submissions and market clearings, State-of-Charge (SoC) constraints, and dynamic price environment. By doing so, we develop a methodology that can be generalised to give insights into the economics of large energy storage capacities in the rapidly evolving energy landscape. We apply the method to a case study of a battery operator in Germany who can trade in the Day-ahead (DA), ID and the Frequency Containment Reserve (FCR) balancing market. Leveraging comprehensive data collected from the German electricity market for the year 2022, we assess the economics of large-scale storage in a period with high price volatility. We provide insights into market participation by analysing bidding strategies that allow for multi-market coordination and answer whether coordination has a monetary advantage.

This paper is structured as follows: Section \ref{sc:literature} reviews advancements in market coordination and SDDP as a solution method for complex decision problems. Section \ref{sc:coordinated_battery_trading} discusses the battery operator's trading problem and data processing in the German electricity market, including market structures and Markov chain estimation. A case study and simulation are presented in Section \ref{sc:case_study}. Our findings are presented in Section \ref{sc:results}, followed by a discussion in Section \ref{sc:discussion}. We conclude in Section \ref{sc:conclusion} with suggestions for further research.

\section{Literature Review}
\label{sc:literature}
This Section delves into the sequential market problem and its implications for coordinated bidding in electricity markets. We examine the evolution of stochastic programming methods for electricity trading under the uncertainty of volatile prices, highlighting the application of SDDP in hydropower reservoir management. We emphasise its effectiveness in dealing with complex multi-stage decisions. Additionally, we discuss various price modelling techniques essential for effective market participation and show our contribution.

\subsection{Sequential Markets and Coordinated Bidding}

Trading in multiple markets simultaneously requires dynamic evaluation of often interrelated price structures across these markets. This complexity necessitates the evaluation of market-specific conditions, pricing trends, and potential cross-market influences. Such complex decision-making requires advanced analytics and extensive data, which might not always be feasible in practice. 

%beginning of coordinated bidding 
The use of stochastic programming methods for coordinated selling of electricity across multiple markets was initially driven by the need for decision support to hedge the risks associated with selling electricity at volatile prices now or in advance on stable yet uncertain future or options markets \citep{arnold_hedging_2002}. Over the years, the setup has been extended to coordinate sequential short-term power markets \citep{triki_optimal_2005, plazas_multimarket_2005, boomsma_bidding_2014, kraft_stochastic_2022}. Early works formulated the problem and qualitatively analysed solutions using small examples. Over time, advancements in computational resources and methodological refinements enabled the implementation of these solutions on more realistic data. Notably, \citet{boomsma_bidding_2014} worked on the problem of a hydropower producer in the Nordics and coordinated bids of the DA and balancing market with sequential dependencies of balancing prices on spot prices. The properties of the balancing market they describe share similarities with the ID market we know today. \citet{ottesen_multi_2018} extend coordination efforts to three sequential markets for selling demand-side flexibility but simplify the decision space by creating new models sequentially each time information is revealed and consolidating the number of stages to three for tractability. As a consequence, limitations on inter-market trading apply. The influence of the gradual revealing of information with the ability to react between decisions becomes apparent with \citet{wozabal_optimal_2020} describing the time gap and complex interplay between markets and stages based on the flow of information. The market coordination problem is further complicated by time-coupling constraints of storage. \citet{lohndorf_value_2022} explicitly model the coordination value at different examples of storages, including grid-connected battery storage, for two spot markets as a major extension for modelling a battery's complicated time coupling constraints. \citet{finnah_integrated_2022} model the coordinated bidding of a pumped hydro storage operator on the two German spot markets by approximating the value function. \citet{heredia_optimal_2018} optimised the operation of a virtual power plant, including battery storage, across the spot and secondary reserve markets using the Iberian electricity zone as a case study.

%beginning of SDDP
Adding markets increases complexity and requires efficient solution methods to calculate optimal actions. Motivated by the complexity of managing hydro reservoirs and their electricity production, SDDP has evolved into the de-facto standard solution technique in complex hydropower reservoir management. SDDP can address multi-stage decisions with time-coupling constraints over a long time horizon \citep{pereira_multi-stage_1991, rotting_stochastic_1992, rouge_using_2016}. \citet{plazas_multimarket_2005} and \citet{fleten_stochastic_2007} applied SDDP to link market coordination in the hydropower sellers' problem under price uncertainty. Since then, many authors have investigated the complicated relationship between price and weather uncertainty by coordinated bidding using stochastic programming and the method has seen gradual methodological improvements. \citet{lohndorf_optimizing_2013} added an exogenous Markov process that allows an approximation of the value function of multiple connected hydro reservoirs. The technique, named approximate dual dynamic programming (ADDP), significantly improved computational performance compared to previous methods.

% Contribution
% What is the value of coordination?
Despite the extensive body of literature, there is no consensus on the monetary benefits of coordination across different markets. Studies by \citet{wozabal_optimal_2020, finnah_integrated_2022, lohndorf_value_2022} find that a value for coordinating bids over markets exists and can be up to 20\,\%. \citet{heredia_optimal_2018} find that coordination of a virtual power plant with a battery on the balancing market can nearly quadruple revenues. In turn, \citet{kongelf_portfolio_2019} find only a small gain from coordination and further describe a dependency on portfolio size. Unwanted incentives for coordination exist, too: \citet{boomsma_bidding_2014} find that under a two-price balancing setup (down-regulation balancing price, upregulation spot market price), it is financially beneficial to hold back capacity under some market conditions by providing down-regulation. 

With the increasing volatility in the market environment and recent price drops of battery technology, the literature lacks models for coordinating spot and reserve markets with the time-coupling constraints of short-term battery storage while obeying the complex decision structure between bidding and execution. We aim to investigate this in the current work. With a three-day planning horizon, we position ourselves at the initialisation step of the operational decisions for short-term energy storage (<4h duration). Furthermore, we coordinate across a total of three markets and use SDDP as a solution algorithm to cope with the curse of dimensionality.

\subsection{Price Modelling}
% check if \citep{lohndorf_modeling_2019} can fit somewhere

When modelling interactions between reserve and spot market prices, accurately representing price movements of the combination of markets is important to train effective trading strategies. A side objective is to achieve the calculations with a reasonable computational effort so that a trader can follow the market schedules in practice. While spot markets and their respective prices have received considerable attention (see, e.g. \citep{ottesen_multi_2018, wozabal_optimal_2020, lohndorf_value_2022}), academic research on balancing market prices, like the FCR, is sparse and point out difficulties in modelling. \citet{baetens_two-stage_2020} highlight the challenges associated with calibrating forecasting models while \citet{backe_predictions_2023} question the significance of predictive information for balancing markets from historical data. Other works on bidding in balancing markets consider market structures different to what is used now in Germany: \citet{klaeboe_benchmarking_2015} benchmark various models for predicting prices and volumes of the Norwegian balancing market, some in combination with spot prices. The authors find that price calibrations are complex and conclude: "[...] the volume and the premium in the balancing market are random. In fact, it could be interpreted as a sign of an efficient electricity market that it is impossible to predict the balancing market price" \citep{klaeboe_benchmarking_2015}. \citet{boomsma_bidding_2014} find strong autocorrelations and cross-correlations between the spot and balancing markets. Specifically, the German balancing market seems to suffer from additional intricacy, as it "is known for hardly explainable prices, supposedly due to a high market concentration" \citep{kraft_stochastic_2022}. Wholesale electricity prices hold significant political and economic importance, yet research on price formation, especially during times of notable price distortions like the German energy crisis is limited \citep{liebensteiner_high_2025}. 

Another important aspect is aligning the resolution and complexity of the price process with the solution algorithm. \citet{finnah_integrated_2022} highlight challenges in solution algorithms that approximate high-dimensional forecasts and state space extensions of dynamic programmes, emphasising the need for careful assessment of the trade-offs between the complexity of the price process and solution quality. Previous works have explored various methods, including Autoregressive Integrated Moving Average (ARIMA) models \citep{klaeboe_benchmarking_2015}, Seasonal Autoregressive Integrated Moving Average (SARIMA) models \citep{boomsma_bidding_2014, baetens_two-stage_2020}, Seasonal Auto-Regressive Integrated Moving Average with exogenous factors (SARIMAX) \citep{kraft_modeling_2020}, neuronal networks \citep{kraft_modeling_2020} and fixed prices approach \citep{henni_industrial_2022}.\footnote{Price modelling success might dependent on the training data. It is important to note that even recent works by the authors of \citet{baetens_two-stage_2020, kraft_modeling_2020} relied on an obsolete German market scheme, which ended on July 1, 2020, and used daily availability auctions instead of the current four-hour interval structure. These market changes further restrict the training data available for calibrating advanced models for this market.} 

\subsection{Contribution}
The paper makes a threefold contribution to the field. It introduces a (1) \textbf{multi-stage stochastic bidding model for battery storage}. We add a balancing service to the DA, and ID spot markets and develop a profitable trading strategy that coordinates bidding while adhering to market schedules. The real-world bidding process is implemented at reduced resolution for computational tractability, maintaining market characteristics. To address the curse of dimensionality, we use SDDP to derive decision policies under uncertainty while being significantly more efficient with training data than machine learning methods. Additional trade-offs of modelling resolution are critically discussed.

% the objective is to show profit creation, coordination among markets and under training time constraints

We construct and calibrate state-of-the-art (2) \textbf{econometric models for price processes}, which we use as a prerequisite for sound decision-making on stochastic markets by our bidding model. These price models are calibrated using data from the challenging market conditions of 2022. We separate the stochasticity from market sentiments, avoiding reliance on exogenous assumptions. This methodological advancement not only provides a framework for future research but also offers valuable insights into the German power markets during the energy crisis in 2022.

%prove that a good representation of stochastic price behaviour is possible from a small amount of data. -> for that we need out of sample testing

The bidding model can be used to (3) \textbf{evaluate battery economics with multiple revenue streams}. While arbitrage operations within or across spot market segments are well-discussed in the literature at length, this paper provides new insights into the operation and trading strategies when adding a balancing market. We compare profits within individual markets under realistic trading behaviour and explore the additional value generated through coordinated operations. Our approach helps determine the value of coordination for a battery over time and across markets, illustrating multi-market business models. While the German market serves as a specific example, the methodology is applicable to short-term markets globally.

\section{Coordinated Battery Trading}
\label{sc:coordinated_battery_trading}
\begin{figure}
    \centering
    \includegraphics[width=0.7\linewidth]{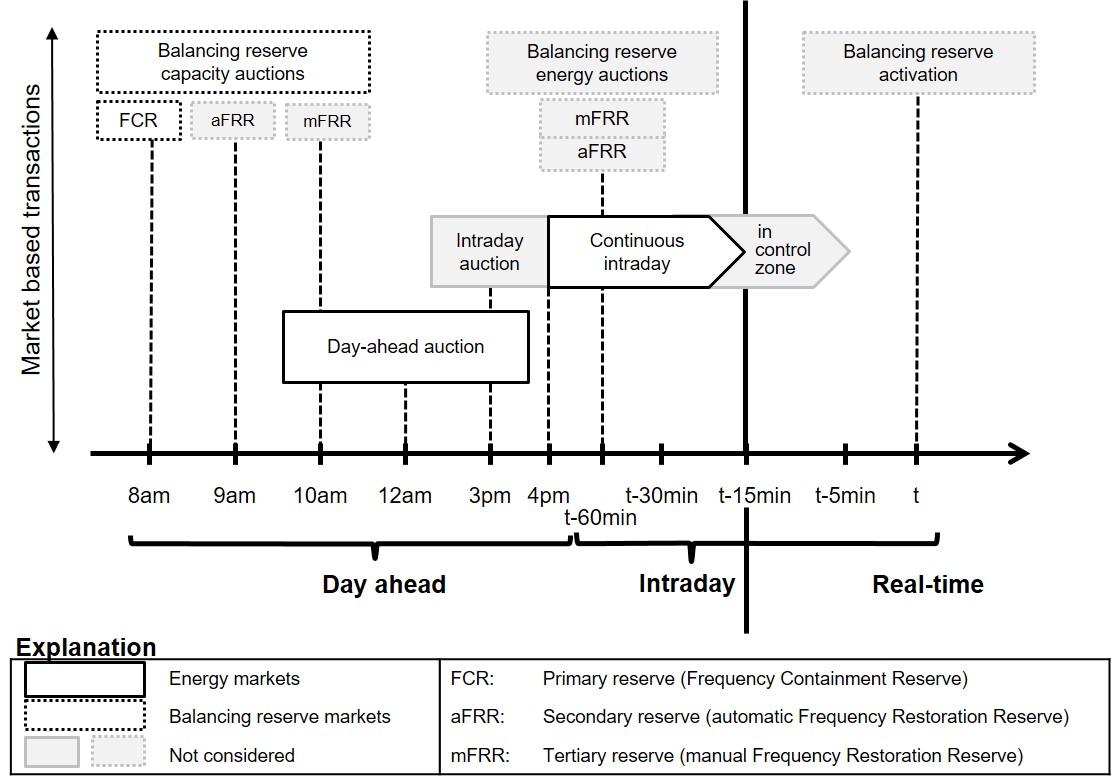}
    \caption{Considered market sequence with reserve markets segments as well as DA and ID spot markets.}
    \label{fig:marketsequence}
\end{figure}

The battery storage operation problem comprises decisions on market participation, bid quantities, price levels, and timing. Unlike traditional generation, batteries do not require long lead times for operation and are highly flexible to react to price changes. There is considerable variation in available information from individual markets bidding to real-time, leading to market volatility that battery storage can economically exploit. Figure \ref{fig:marketsequence} illustrates the schedule of electricity markets. Power reservations can be made in the FCR balancing market before spot markets close. The DA market, cleared daily between 1 p.m. and 2 p.m., provides price-quantity pairs for the next day. Market participants must adjust their DA positions based on updated information, mainly weather and RES share of total generation \citep{kiesel_econometric_2017}, with corrective trading in the ID market until 30 minutes before delivery.

The complexity of the problem stems from its multidimensional nature, driven by the interplay between the markets, time coupling constraints of the battery, and the multi-day optimisation horizon. To cope with the market sequence's complexity, traders can (and do in practice \citep{heredia_optimal_2018}) simplify decision-making by (1) focusing only on a subset of markets, (2) making sequential decisions following the market schedule, (3) limiting the foresight and planning horizon, (4) neglecting information about uncertainty or (5) expanding the market intervals to fewer decision periods. 
%Our trading model builds on three coordinated markets on a three-day planning horizon does not decompose the problem sequentially, considers a realistic representation of uncertainty and only shortens the market's intervals to four hours.

\subsection{Market Structures}

\begin{figure}[htbp]
    \centering
    \includegraphics[width=0.9\columnwidth]{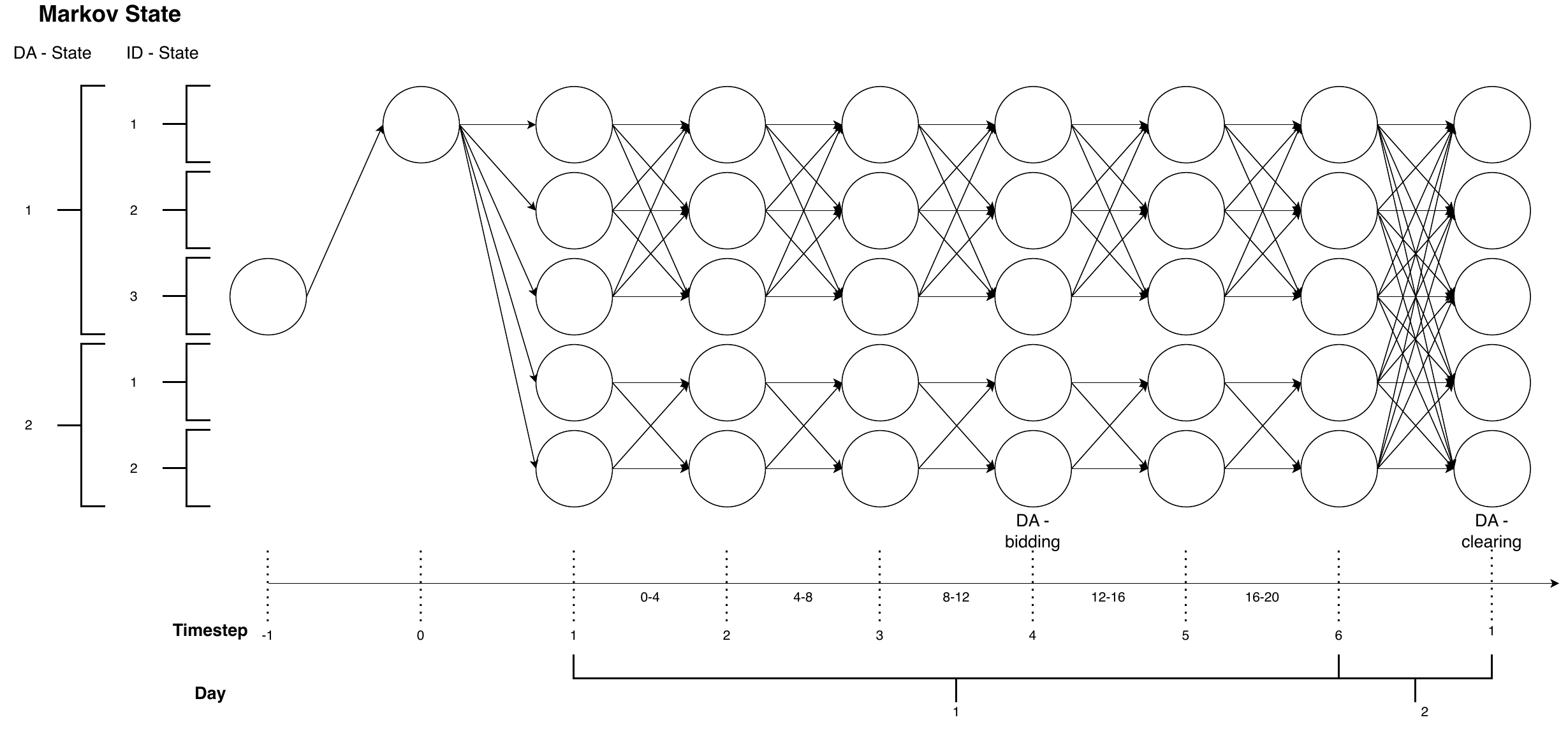}
    \caption{Exemplary visualisation of the stochastic process for the two markets of ID and DA}
    \label{fig:SDDP_process}
\end{figure}

This Section describes the relationship between individual markets and the fundamental model assumptions. We use a Markov Chain to model prices, where a finite number of states are interconnected by conditional probabilistic movements called state transitions \citep{kirkwood_markov_2015}. A state is a discrete point in time that contains available information (prices in our case). Only one state can be visited at a time, and the process is memoryless, meaning that the transitions depend solely on the currently active state. %Figure \ref{fig:SDDP_process} visualizes such a Markov Chain.

%The model's time resolution should be able to capture the most crucial decision points and information revelations. 
Our model emphasises the correct timing of the German electricity market schedule but can easily be adapted to other use cases. The lowest common denominator of all markets is the four-hour resolution of the FCR market. Sub-stages for both other markets are possible but not considered in this work. A 24-hour operational day $d$ is divided into six four-hourly time blocks $f \in \left\{ 1, ..., 6\right\}$, where $f = 1$ represents the interval 00:00–04:00, $f = 2$ the interval 04:00–08:00, and so on. A combination $(d,f)$ defines a stage $t$ of the decision problem. The model starts at midnight, with no commitments for the DA and FCR markets before their first clearing since we do not consider a possible deterministic pre-clearing from the previous day. We do not allow bidding on the FCR and DA markets on the last day of the planning horizon to avoid running into end-of-horizon distortions. The individual markets $m \in \{DA, ID, FCR \}$ are modelled as follows:
\begin{enumerate}
    \item \textbf{FCR Market}: At day $d-1$, in block $f=3$, the FCR market takes capacity bids on a four-hour resolution for the next day $d$. The market is then cleared at $t=(d-1,4)$. Bidding and clearing are done for the whole next day in all six four-hour blocks. 
    
    \item \textbf{DA Market}: At day $d-1$, in block $f=4$, we bid for all six blocks of day $d$. We defer the market clearing to $f = 1$ the next day $d$, again for clearing of the whole next day. \footnote{This modelling choice is motivated by the dependency of ID prices on DA prices; ID price selection requires the current DA state. An early clearing (for the next day) would overwrite the DA state needed for ID prices of the current day. Unlike variable values, we can not temporarily store Markov chain movements. By extending the clearing to midnight, we preserve the location of the graph at the cost of not revealing the cleared quantities two steps before the next day.} 
    
    %Alternatives have been considered and include extending the state space of the Markovian policy graph by one dimension and a DA price- and state-independent ID price process. The first comes at much higher complexity and computational costs, whereas the latter loses the price relationship between the two markets and the autocorrelative behaviour. \\
    
    %Price developments in this market follow the same Markov chain approach with multivariate price paths for the subsequent six steps. 
    \item \textbf{ID Market}: The ID market price is modelled as a DA price-dependent spread. For a block $f-1$, bidding occurs for the next block $f$, with the commitment delivered in $f$. The market has three levels: one at the mean of the distribution against DA prices, and one level each above and below this price, similar to the discrete intra-stage price process in \citep{lohndorf_optimizing_2013}.
    
    %\textcolor{orange}{actually, I think this is a good argument to keep the ID randomness out of the Markovian Process and make it independent. State transitions are always independent of the previous ID state, as well as independent of the other markets. Similar to the stage-independent noise in the SDDP.jl examples (the sub markovian noise)}
\end{enumerate}

Figure \ref{fig:SDDP_process} shows a simplified Markov chain with transitions for two of the three markets. The DA stochastic process changes states from bidding in $f = 4$ in $d-1$, revealing the uncertainty in $f=1$ in $d$. The implied cost of an operator's decisions may enter the objective function with a delay between bidding and clearing, which makes optimal decision-making more complex. 
%We assume no inherited DA and FCR commitments before the first clearing. Until then, only ID market actions are allowed on the first day. 

\subsection{Price Modelling}
We argue that due to recent electricity price distortions, simple time series analysis is inadequate for capturing price movements when exposed to external shocks. Training models on ordinary price years may result in a poor fit during exceptional years. To address this, we normalise the data by separating prices into a predictable component based on fundamental models that are shown to provide accurate estimations in the past and can capture non-linearities \citep{pape_are_2016} and a stochastic component. 
%The stochastic price movements, relevant for short-term market price changes, can predominantly be explained by variations in weather, load, plant unavailability and market conditions. In practice, battery operators might consult commercial forecasts with prediction tools or specialised companies for macroeconomic regressors while keeping the relative scenario generation in-house. 

\begin{figure}[htbp]
    \centering
        \includegraphics[width=0.8\textwidth]{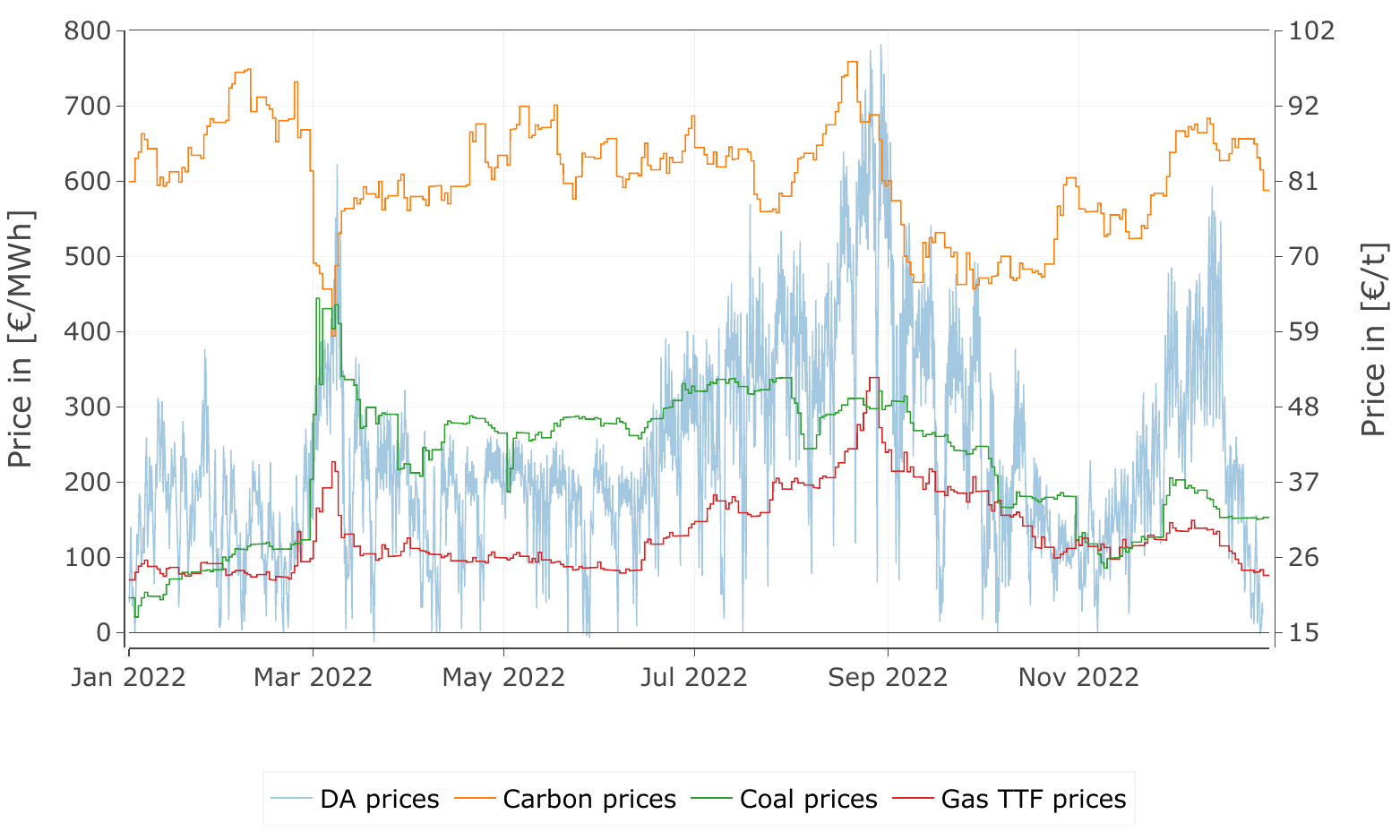}
        \caption{Econometric parameters and DA prices in the year 2022}
        \label{fig:fundamentals}
\end{figure}

We use the relative stochastic component to create relative price scenarios, which we combine with forecasts of fundamental model parameters for situation adjustment of the bidding, which are then used to estimate the Markov chain of price transitions. The three markets of DA, ID and FCR market prices observed a high price level with increased volatility throughout the year 2022 (Figure \ref{fig:prices_2022}), especially compared to 2021 (Figure \ref{fig:prices_2021}). We selected residual load, Gas Title Transfer Facility (TTF), and $CO_2$ price as explanatory variables based on their statistical significance and the resulting adjusted $R^2$. Figure \ref{fig:fundamentals} depicts the evolution of the fundamental explanatory variables throughout the year. 

In a next step, we reduce the DA and ID time series from hourly to 4-hour resolution by calculating the mean value within each interval. We then split the DA and FCR price time series into a deterministic and a stochastic component by Ordinary Least Squares (OLS) econometric models. Further, we calculate the $10\,\%$ lower- and upper quantiles of residual load for later usage as explanatory variables in the econometric price separation. 
%In Section \ref{sec:estimating_markov_chain}, we use the stochastic component for estimating the Markov chain.
%Table~\ref{tab:econometric_variables} provides descriptions and notation for the explanatory variables used in the subsequent sections. 

%\begin{table}[htbp!]
%    \centering
%    \caption{Variables for the econometric models used and their variable type}
%    \label{tab:econometric_variables}
%    \begin{tabular}{lll}
%        \toprule
%        Variable & Content & Variable type \\
%        \midrule
%         $X_1$ &  Gas TTF price & continuous\\
%         $X_2$ &  Carbon price  & continuous \\
%         $X_3$ &  Residual Load & continuous \\
%         $X_4$ &  Upper quantile residual Load & dummy \\
%         $X_5$ &  Lower quantile residual load & dummy\\
%         $X_6$ &  Weekend & dummy \\
%         $X_7$ &  $f=1$ of the day & dummy \\
%         $X_8$ &  $f=2$ of the day& dummy \\
%         $X_9$ &  $f=3$ of the day & dummy \\
%         $X_{10}$ & $f=4$ of the day& dummy \\
%         $X_{11}$ & $f=5$ of the day & dummy \\
%        \bottomrule
%    \end{tabular}
%\end{table}

\paragraph{DA Prices}
\label{sec:DA_prices}

The DA price $P^{DA}_{t}$ from the historical time series can be separated into a deterministic term $D^{DA}_{t}$ and stochastic residual $S^{DA}_{t}$.
%\begin{align}
%    P^{DA}_{t}  &= D^{DA}_{t} + S^{DA}_{t}. \\
%    P^{DA}_{t}  &= \beta_0 + \beta_1 X_1 + \beta_2 X_2 + \beta_3 X_3 X_4 + \beta_4 X_3 X_5 + \beta_5 X_6 + S^{DA}_{t}.
%\end{align} 
Explanatory variables are the Dutch TTF gas price, Carbon certificate price and the residual load. The model achieves an adjusted $R^2$ of 84.8\,\% in 2022 and 84.9\,\% in 2021. Therefore, it is deemed a good fit to predict price developments with few explanatory variables. We find strong positive autocorrelations with a Durbin-Watson value of 0.154.

\paragraph{ID Prices}

ID prices can be modelled as a price spread, being dependent on previous DA price realizations \citep{hagemann_price_2013, rintamaki_strategic_2020}, or as an independent equilibrium between ID supply and demand \citep{pape_are_2016}. We use the first approach and model ID prices as up- or downward spreads of forecasting errors from the previously cleared DA price where $P^{ID} = P^{DA} + ID^\text{spread}$.

\paragraph{FCR Prices}

%Our literature review shows that the FCR market price is particularly challenging to model. However, our trading policy depends on reliable price scenarios that contain meaningful information rather than just noise. This is important to ensure that our trading strategy is not adversely affected. 
We consider the FCR market to be price-independent of other markets and estimate it by another linear regression similar to the DA price model. We find hardly a linear correlation between DA and FCR market prices (correlation coefficient of -$0.024$), and the time of the clearing is before the clearing of the DA market. It might be possible to anticipate and behave strategically, but we do not find evidence from the time series to support this assumption.
%\begin{equation}
%    P^{FCR}_{t} = D^{FCR}_{t} + S^{FCR}_{t}.
%\end{equation} 
We use dummy variables for the day's different four h intervals $f$ to capture time-dependent patterns. Furthermore, we use the log function for the prediction and residual demand quantiles as independent variables to include scarcity effects.
%\begin{equation}
%    \begin{aligned}
%        \log(D^{FCR}_{t}) &= \beta_0 + \beta_1 X_1 + \beta_3 \log(X_3) X_4 + \beta_4 \log(X_3) X_5 + \beta_5 X_6 + \beta_6 X_7 \\
%        &+ \beta_7 X_8 +\beta_8 X_9 + \beta_9 X_{10} + \beta_{10} X_{11}.
%    \end{aligned}
%\end{equation}
Including residual demand quantiles as a regressor significantly improves predictability. Residual demand represents the demand not satisfied by renewable sources like wind, solar, and hydropower. In 2022, this approach increased the explained variance in the OLS model to an R-squared value of 36\,\%. In 2021, the same model achieved an R-squared value over 53\,\%. We find no significant signs of autocorrelation in the FCR prices. For further information on how to estimate quantile levels, we refer to the work of \citet{aneiros_functional_2013, do_residual_2021}, who achieved good results in short- and long-term predictions of residual demand using a functional nonparametric model and quantile regression respectively.

%Estimating the quantile levels in advance is not within the scope of this work, and we assume perfect knowledge of the correct levels. Simple interpolations on our dataset with residual demand from the year 2021 to 2022 show that correcting the annual sum of demand and renewable share results in an $\sim 8\,\%$ overestimation of the lower quantile and an $\sim 3.4\,\%$ underestimation of the upper quantile. This shows that simple methods are sufficient to support our assumption in practice. 

\subsection{Estimating the Markov Chain}
\label{sec:estimating_markov_chain}
\begin{figure}[htbp]
        \centering
        \includegraphics[width=\textwidth]{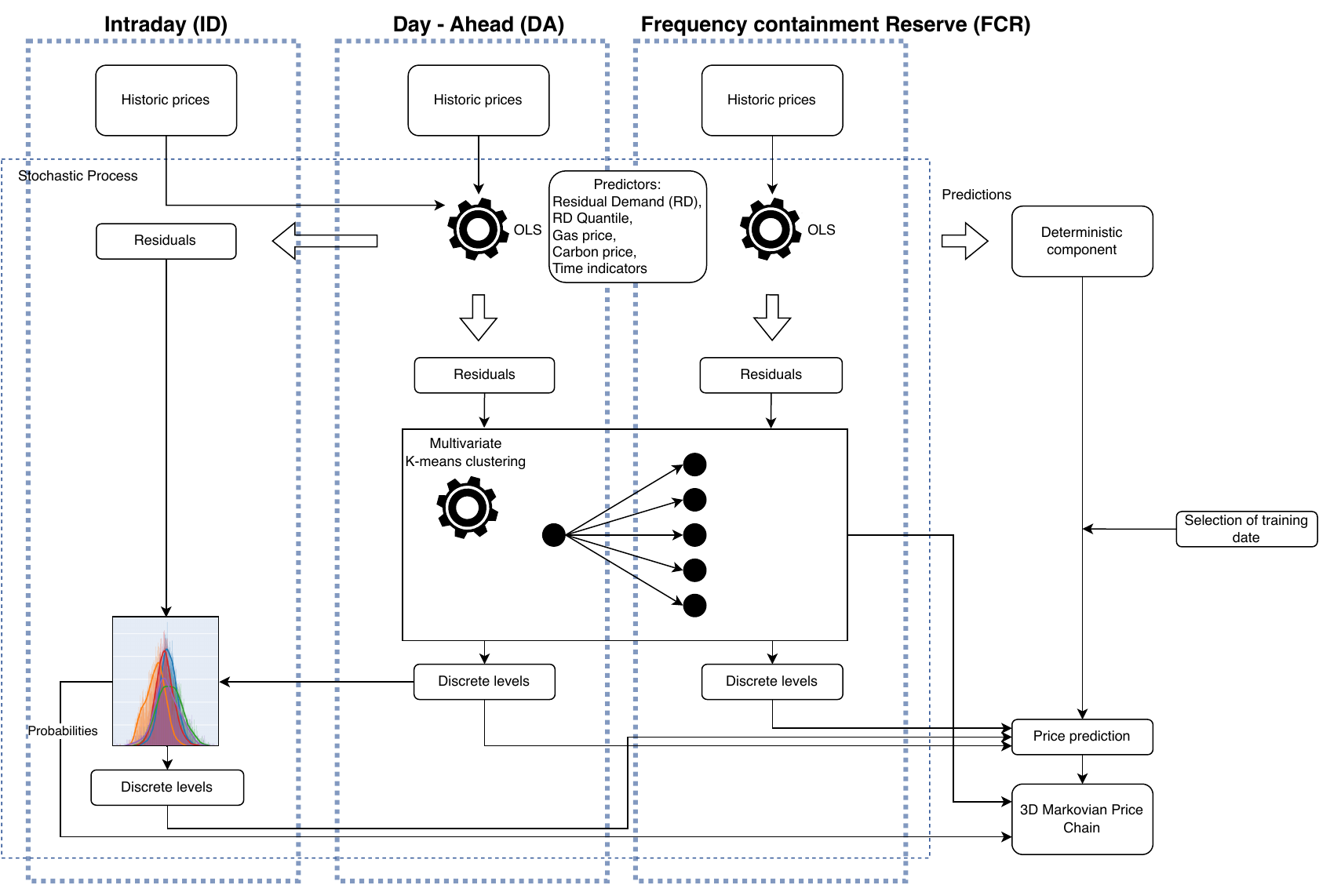}
        \caption{Flow chart of processing for estimating the Markov chain}
        \label{fig:preprocessing}
\end{figure}

%The following explains how we get data from historical time series to input into our SDDP model. 
The model requires discrete price states for the individual markets and a corresponding transition probability lattice to define the Markov chain. Figure \ref{fig:preprocessing} provides a schematic overview of the individual steps involved.
%\begin{figure}[htbp]
%        \centering
%        \includegraphics[width=0.3\textwidth]{figs/ID_distplot_example_bounds.pdf}
%        \caption{ID spread}
%\end{figure}

\begin{figure}[htbp]
    \centering
    \includegraphics[width=0.7\columnwidth]{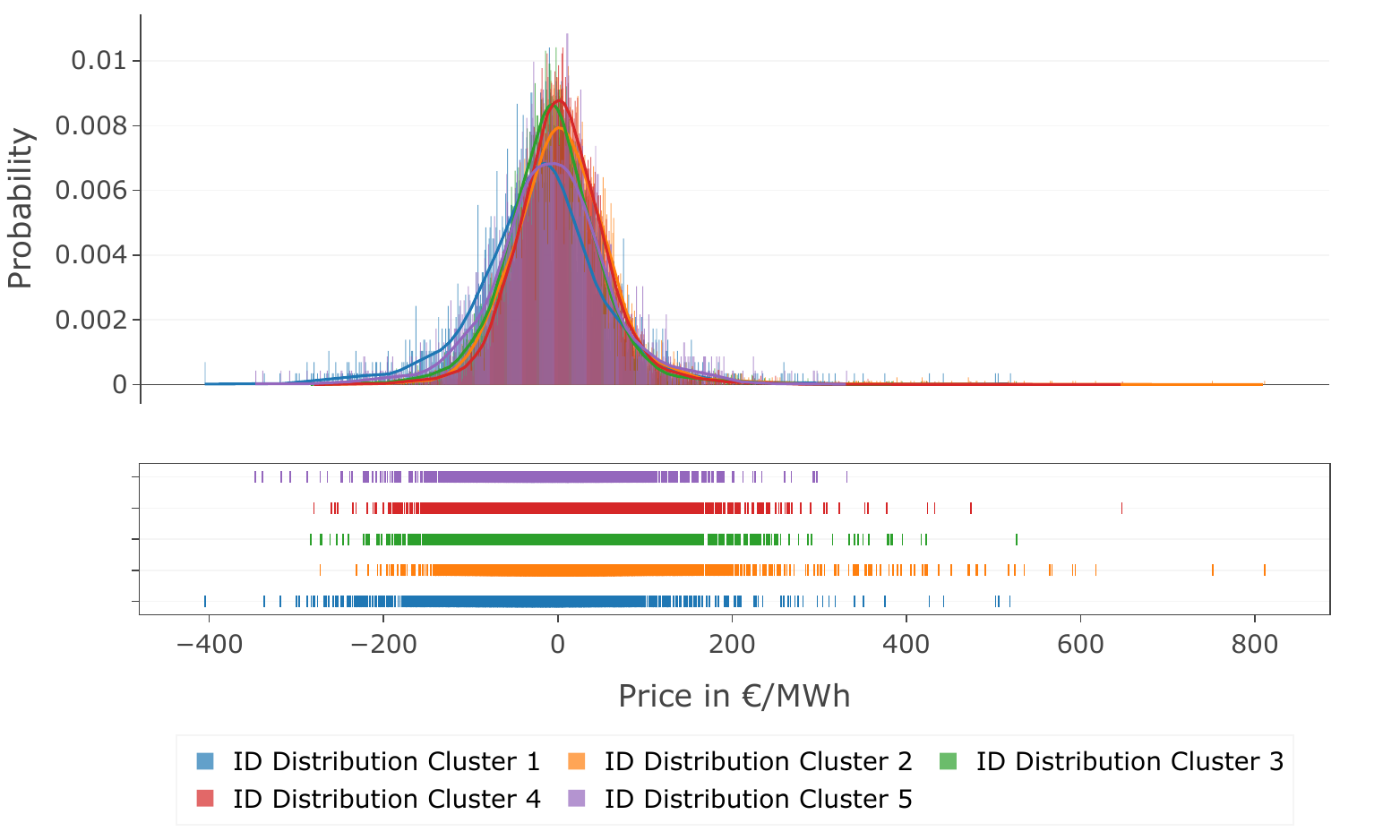}
    \caption{Histogram of the distances from the stochastic ID realisations to the centroids of the DA clusters of the year 2022 with dependency on DA cluster}
    \label{fig:ID_probability_density}
\end{figure}

%Price levels
We apply multivariate Euclidean k-means clustering to the residuals of the stochastic price components. Our analysis uses a sequence of three consecutive days, resulting in a total of 121 historic price combinations (363/3). Based on the elbow plots, we determine that the optimal number of clusters is 3 for the FCR market and 5 for the DA market. The results of this clustering analysis are presented in Figures in \ref{sec:clustering}.
%The residuals of the econometric models are input for the clustering of stochastic price components. Three successive days form a sequence for the later SDDP model, totalling 121 (=363/3) historic price combinations. The last two days of the year are omitted to make the allocation integral. The FCR and DA markets require a prediction for six consecutive 4-hour intervals of the next day. We reduce these 121 sequential price movements by employing a multivariate Euclidean k-means clustering approach. We then achieve a reduced number of representative clusters, consisting of six consecutive four-hour data points, for each of the three days of the planning horizon. Figure \ref{fig:DA_elbow} presents elbow plots to determine the number of necessary clusters. We observe two slight elbows at a cluster count of 3 and 5 clusters. Based on this, we proceed with five clusters as discrete descriptions of stochastic price levels. The same procedure is applied to the FCR market, with clusters depicted in Figure \ref{fig:FCR_cluster}. We decide on three clusters based on the elbow plot in Figure \ref{fig:FCR_elbow}. 

Figure \ref{fig:ID_probability_density} shows the probability density functions of ID deviations for the different DA clusters. The probability density function is characterised by long tails of deviations in both directions, particularly in the positive price direction. Tests for normality of the ID deviations were rejected in all DA clusters despite desirability from an efficient market theory perspective. We discovered that the distribution of ID spreads differs in mean and variance depending on the DA price cluster. We take advantage of this and derive the ID price as a cluster-dependent difference to DA prices with $P^{ID}_{clusterDA} = P^{ID}_{historical}-P^{DA}_{clusterDA}$.
Based on the distributions of the differences, we define three discrete ID price levels for each of the five DA levels and the respective probabilities: the mean and the 15 and 85 percentiles of each ID distribution.

%lattice calculation
Next, we calculate the transition probabilities between Markov states based on the discrete cluster allocation of historical prices and their changes over time.% Once we have the cluster allocation for a sequence of days, we can analyse the allocation of the preceding and succeeding sequences to calculate the transition probabilities. This is done by counting the transitions and weighting the counts appropriately based on the cluster allocation.
In the last step, we built a prediction for the investigated period shortly before market participation. We use the dependent variables of the OLS model and assume that market participants have access to in-house- or external forecasts. The forecasted fundamentals and clustered stochastic residuals are then merged into a Markov chain, representing the current market situation with stochastic market uncertainty. Based on these prices, we solve the SDDP model and approximate the optimal trading policy for the battery operator. 

A comparison of the discrete prices in the Markov chain (green) to the historical prices (blue) in an exemplary period is visualised in \ref{ch:scenarios_investigated_period}. 

\subsection{Mathematical Model} \label{sec:math_model}
In this section, we explain our coordinated multi-market battery storage trading model. Table \ref{tab:nomenclature} lists parameters, variables and sets. 
\begin{longtable}{p{0.2\textwidth}p{0.75\textwidth} ll}\label{tab:nomenclature}\\
    \caption{Designated sets, parameters, and variables of the mathematical framework.}\\
    \toprule
		\multicolumn{2}{l}{\textbf{Sets}}  \\ 
        $\mathcal{M}$ & Set of markets: $m \in \left\{DA,ID, FCR\right\}$   \\
        $\mathcal{N}^m$ & Set of price levels: $n_m$  \\
        $\mathcal{D}$ & Set of days: $d$  \\
        $\mathcal{F}$ & Set of 4 hour blocks within a day: $f$ \\
        $\mathcal{T} =  \mathcal{D} \times \mathcal{F}$  & Set of time stages indexed by $t=(d,f)$\\
        \midrule
        \multicolumn{2}{l}{\textbf{Parameters}}  \\ 
        ${Q}$        & Storage capacity of the battery \\
        ${Q^{{\tt start}}}$& Start- and end storage level of the battery \\
        ${L}$        & Rated power of the battery\\
        %${E}$        & Single trip efficiency of the battery\\
        ${\rho}$        & Penalty term for the usage of slack variables\\
        %${\eta}$    & Round trip efficiency factor \\
        \midrule
        \multicolumn{2}{l}{\textbf{Random Variables}} \\
        $P^m_{tf}$ & Cleared market price in market $m$ at time stage $t$ in four hour block $f$ [\euro{}/MW] \\
        \midrule
        \multicolumn{2}{l}{\textbf{State Variables}} \\
        $SoC_{t}$ & State of Charge in time stage $t$ [MWh]\\
        ${x}^m_{tfn}$ & Bid quantities in market $m$, at price level $n$ for block $f$ in time stage $t$ [MWh]\\
        $\tilde{y}^m_{tf}$ & Offset corrected committed market quantities for market $m$ at time stage $t$ for block $f$ [MWh] \\
        $z^{m}_{tf}$ & Helper variable for offset correction at time stage $t$ and for block $f$ [MWh]\\
        \midrule
        \multicolumn{2}{l}{\textbf{Local Variables}}\\
        $y^m_{tf}$ & Market clearing quantities at the time of clearing for market $m$ and block $f$ [MWh] \\
        $s_{t}$ & Slack variable \\
        \bottomrule
\end{longtable}

\subsubsection{Objective Function}
The objective function maximises the profit from trading on the FCR, DA and ID markets by summing the products of prices and quantities. The FCR and DA markets clear daily in periods $f=4$ and $f=1$, respectively. We add a penalty term for storage level violations.  
\begin{equation}
    \begin{aligned}
    \mathrm{max} \quad 
    &\mathbbm{1}_{\left\{4\right\}}(f') \cdot \sum_{f\in F} P^{FCR}_{tf} y^{FCR}_{tf} +
    \mathbbm{1}_{\left\{1\right\}}(f') \cdot \sum_{f\in F} P^{DA}_{tf} y^{DA}_{tf} \label{eq:objectivesingle} \\
    &+ P^{ID}_t y_{f}^{ID} + 
    \rho s_t,
    &&\quad t = (d,f') \in \mathcal{T}.
    \end{aligned}
\end{equation}

\subsubsection{Constraints}

Equation \eqref{eq:marketclearing} is the market clearing constraint. Each market is cleared according to quantity bids on discrete price levels, matching the discrete levels in the Markov states. The step-function $\mathbbm{1}_{(-\infty, P^m_{tf})} (S^m_{tfn})$ returns 1 if the price of the bid is lower than or equal to the sampled market price ($S \leq P$) and 0 if higher ($S > P$). 
\begin{equation}
    y^m_{f} = \sum_{n \in \mathcal{N}_m} \mathbbm{1}_{(-\infty, P^m_{tf})} (P^m_{tfn}) (x^m_{t-1,f,n}-x^m_{t-1,f,n-1}), \quad  f \in \mathcal{F}, t\in \mathcal{T}_m^{\tt clearing}, m \in \mathcal{M}. \label{eq:marketclearing} 
\end{equation}

Bids for the ID and FCR markets are submitted one timestep before respective market clearing, with the DA market submitting bids in block four each day. We follow the approach of \citet{fleten_constructing_2005}, \citet{lohndorf_optimizing_2013}, and \citet{wozabal_optimal_2020} to model the bidding function: quantity bids are implemented on the discrete price levels as monotonically increasing bid curves, as ensured by Constraint \ref{eq:monotonically_increasing}. 
\begin{equation}
     x_{tf(n-1)}^m \leq x_{tfn}^m, \quad  n \in \mathcal{N}, f \in \mathcal{F}, t \in \mathcal{T}_m^{\tt clearing}, m \in \mathcal{M}. \label{eq:monotonically_increasing}
    %x_{tf(n)}^{ID} &\leq SoC_{t+1} - \tilde{y}_{(t+1),f}^{DA} - \tilde{y}_{(t+1),f}^{FCR} && \forall n \in \left\{\mathcal{\overline{N}}\right\}, f \in \mathcal{F}, t \in \mathcal{T}_m^{\tt clearing},  m \in \mathcal{ID}\\
    %x_{tf(n)}^{ID} &\geq SoC_{t+1} - Q - \tilde{y}_{(t+1),f}^{DA} - \tilde{y}_{(t+1),f}^{FCR} && \forall n \in \left\{1\right\}, f \in \mathcal{F}, t \in \mathcal{T}_m^{\tt clearing}, m \in \mathcal{ID}\\
    % old x_{tf(1)}^{ID} &\geq - SoC_{t+1}  - \tilde{y}_{(t+1),f}^{DA} + \tilde{y}_{(t+1),f}^{FCR} && \\
\end{equation}

\textbf{ID constraints}
We limit ID bidding quantities to quantities not exceeding storage constraints, including previously cleared markets. This is not expected to limit the quality of the solution since it only excludes non-optimal parts of the solution space. Special care has to be taken to the last period each day  $\mathcal{T}_{ID}^{{\tt rest}}=\left\{ (d,f): d\in\mathcal{D}, f=6\right\}$ where the restriction apply to bids, instead of cleared quantities. We have verified the validity of this assumption in \ref{ch:without_ID_const}. 
{\small
\begin{align}
    x_{tfn}^{ID} &\leq SoC_{t+1} - \tilde{y}_{(t+1),f}^{DA} - \tilde{y}_{(t+1),f}^{FCR}, && n \in \left\{\mathcal{\overline{N}}\right\}, f \in \mathcal{F}, t \in \mathcal{T} \setminus \mathcal{T}_{ID}^{{\tt rest}},  m \in \mathcal{ID}. \label{eq:ID_constraint_beginning}\\
    x_{tfn}^{ID} &\geq SoC_{t+1} - Q - \tilde{y}_{(t+1),f}^{DA} - \tilde{y}_{(t+1),f}^{FCR}, && n \in \left\{1\right\}, f \in \mathcal{F}, t \in \mathcal{T} \setminus \mathcal{T}_{ID}^{{\tt rest}}, m \in \mathcal{ID}.\\
    x_{tfn}^{ID} &\leq SoC_{t+1} - \tilde{x}_{(t+1),f,n}^{DA} - \tilde{y}_{(t+1),f}^{FCR}, &&  n \in \left\{\mathcal{\overline{N}}\right\}, f \in \mathcal{F}, t \in \mathcal{T}_{ID}^{{\tt rest}},  m \in \mathcal{ID}.\\
    x_{tfn}^{ID} &\geq SoC_{t+1} - Q - \tilde{x}_{(t+1),f,n}^{DA} - \tilde{y}_{(t+1),f}^{FCR}, &&  n \in \left\{1\right\}, f \in \mathcal{F}, t \in \mathcal{T}_{ID}^{{\tt rest}}, m \in \mathcal{ID}.
\end{align}
}

\textbf{Offset constraints} The offset between market clearing timing and real-time deliveries requires caching of commitments in additional variables for the FCR market. Otherwise, the cleared quantities for the rest of the day would be overwritten with new quantities. We ensure correct values by caching variables: At the time of clearing, commitments for the first part of the next day (until the next clearing) are updated as described in Equation \eqref{eq:offset1}. The commitments for the second part of the day are cached as in Equation \eqref{eq:offset4}. At the start of a new day, Equation \eqref{eq:offset2} then inserts cached commitments into the actual commitments. The information is passed on for all other periods, as given in Equation \eqref{eq:offset3} and \eqref{eq:offset5}.

\begin{subnumcases}{\tilde{y}^{m}_{tf} =} %subnumcases
    y^{m}_{tf}, & $ \quad  f\in \mathcal{F} \setminus \mathcal{F}_m^{{\tt cache}}, t \in \mathcal{T}_m^{{\tt clearing}}, m \in \left\{ FCR\right\},$ \label{eq:offset1} \\
    z^{m}_{tf}, & $ \quad    f \in  \mathcal{F}_m^{{\tt cache}}, t \in \mathcal{D} \times \left\{ 1\right\}, m \in \left\{ FCR\right\}, $ \label{eq:offset2} \\
    \tilde{y}^{m}_{(t-1)f}, &\quad    {otherwise}. \label{eq:offset3} 
\end{subnumcases}

\begin{subnumcases}{z^{m}_{tf} =} %subnumcases
    y^{m}_{f}, & $ \quad    f\in \mathcal{F}_m^{{\tt cache}}, t \in \mathcal{T}_m^{{\tt clearing}}, m \in \left\{ FCR\right\},$ \label{eq:offset4} \\
    z^{m}_{(t-1)f}, &  \quad    {otherwise}. \label{eq:offset5} 
\end{subnumcases}

%No DA caching anymore
We have that $\mathcal{F}_{FCR}^{{\tt cache}} = \left\{ 4,5,6\right\}$, $\mathcal{T}_{FCR}^{{\tt clearing}} = \left\{ (d,f): d\in\mathcal{D}, f=4\right\}$, and $\times$ is used to denote the Cartesian product of two sets. %The DA clearing set is denoted as $\mathcal{T}_{DA}^{{\tt clearing}} = \left\{ (d,f): d\in\mathcal{D}, f=1\right\}$.  %$\mathcal{F}_{DA}^{{\tt cache}} = \left\{5,6\right\}$ and 

The \textbf{state of charge} should always stay within boundaries set by the capacity of the battery, reduced by capacity reservations for the FCR, $\tilde{y}_{tf}^{FCR}$. We assume that the reserved power in the FCR market is available in both directions, and we need to ensure that the respective power is covered by an appropriate SoC level. Hence, we reserve the respective up and down capacities. Slack variables, which are penalised in the objective, relax these constraints and ensure the SDDP algorithm's feasibility. We assume no efficiency losses for the battery since our focus is on a short-term operation where the high-efficiency rates of commercial battery racks are considered neglectable for operational decisions. Furthermore, the problem becomes more tractable.
%Deviations from the storage boundaries are penalised using the slack variables. Equation \ref{eq:absolute1} and \ref{eq:absolute2} describe the absolutes of storage level influencing variables by battery charging or discharging. To keep the problem linear and bring it into standard form, we introduce the variable $u$. 
\begin{subequations}
\begin{align}
    {SoC}_t &= {SoC}_{t-1} - (y^{ID}+\tilde{y}^{DA}_{tf}),%-0.05*u_t
    & &  t \in \mathcal{T}.\\
    %y^{ID}+\tilde{y}^{DA}_{tf} &\leq u_t & & \quad \forall t \in \mathcal{T} \label{eq:absolute1}\\
    %- (y^{ID}+\tilde{y}^{DA}_{tf}) &\leq u_t  & & \quad \forall t \in \mathcal{T} \label{eq:absolute2}\\
    \tilde{y}_{tf}^{FCR} - s_t  &\leq {SoC}_{t} \leq Q - \tilde{y}_{tf}^{FCR} + s_t, && t \in \mathcal{T}. \label{eq:SoC_up_down_constraint}\\ 
    {SoC}_t+s_t &= Q^{INIT}, && t \in \left\{ 0, \overline{T}\right\}. \
\end{align}
\end{subequations}
%just an experiment, delete once falsified -> problem with the penalty of u
%\begin{subequations}
%\begin{align}
%    {SoC}_t &= {SoC}_{t-1} - u_t& &\quad  \forall t \in \mathcal{T}\\
%    \eta \cdot (y^{ID}+\tilde{y}^{DA}_{tf}) &\leq u_t & & \quad \forall t \in \mathcal{T} \label{eq:absolute1}\\
%    - (y^{ID}+\tilde{y}^{DA}_{tf}) &\leq u_t  & & \quad \forall t \in \mathcal{ T} \label{eq:absolute2}\\
%    \tilde{y}_{tf}^{FCR} -s^{down}_t  &\leq {SoC}_{t} \leq Q - \tilde{y}_{tf}^{FCR} + s^{up}_t, &&\quad \forall t \in \mathcal{T} \label{eq:SoC_up_down_constraint}\\ 
%    {SoC}_t+s_t^{up}-s_t^{down} &= Q^{INIT} &&\quad \forall t \in \left\{ 0, \overline{T}\right\} \
%\end{align}
%\end{subequations}

\textbf{Domains}. Trading of the battery in all markets and for all variable types is restricted by the battery's rated power and storage capacity. Bidding volumes in the ID and DA markets are limited by the battery's rated power, while the FCR bids are limited to half the battery's rated power as a conservative assumption. SoC is limited by the battery's capacity, and slack variables are not limited. %Variable $u_t$ is restricted by a combined volume of ID and DA market commitments that is still feasible without violating the storage capacity. %\textcolor{red}{u is limited to 2L because the tight bounds of L cause infeasibilities that would need a relaxation with slack variables, like the storage bounds constraints in \ref{eq:SoC_up_down_constraint}.}
\begin{align}
     - L &\leq y^m_{t,f} \leq L,           &\quad & t\in \mathcal{T}, f \in \mathcal{F}, m \in \mathcal{M}.\\
     - L &\leq z^m_{t,f} \leq L,           &\quad & t\in \mathcal{T}, f \in \mathcal{F}, m \in \mathcal{M}.\\
     - L &\leq\tilde{y}^m_{t,f} \leq L,    &\quad & t\in \mathcal{T}, f \in \mathcal{F}, m \in \mathcal{M}.\\
     - L &\leq x^m_{t,f,n} \leq L,         &\quad & t\in \mathcal{T}, f \in \mathcal{F}, n\in \mathcal{N}^m, m \in \left\{DA, ID \right\}.\\
     0 &\leq x^m_{t,f,n} \leq 0.5 L,         &\quad & t\in \mathcal{T}, f \in \mathcal{F}, n\in \mathcal{N}^m, m \in \left\{FCR\right\}.\\
     0 &\leq SoC_t\leq Q,                  &\quad & t\in \mathcal{T}\setminus \left\{ 0, \overline{T}\right\}.\\
     s_t &\in \mathcal{R},  &\quad & t\in \mathcal{T}.
     %-2L &\leq u_t \leq 2L &\quad &\forall t\in \mathcal{T}
\end{align}
%\textbf{Choice of Parameters}

%Storage violations at the end of the time horizon or within the investigated time span are penalised with the same factor of $\rho$. We compared $\rho = 100,000\euro{}/MWh$ and $\rho = 10,000\euro{}/MWh$ $\rho = 3,000\euro{}/MWh$ and found that the smallest penalty is sufficient to train a reasonable policy. \\

% excluded. We had major problems with convergence and penalising the magnitude of net storage action losses. 
%The round trip efficiency is estimated at 0.9 \citep{barbry_robust_2019}, which is a conservative choice compared to other publications in the field \citet{lohndorf_value_2022}.  

\section{Case Study and Implementation}
\label{sc:case_study}
We have implemented a case study on a 10MW/10MWh battery storage located in the German electricity market zone and used data from EpexSpot for the year 2022. While spot markets have not seen many regulatory changes in recent years, the FCR market structure changes frequently. At the time of writing, reservations are cleared in a pay-as-cleared remuneration system. The market is intended for small imbalances and therefore procured as a symmetrical product with at least 1 MW power and 30 seconds of activation time.\footnote{\href{https://www.regelleistung.net/en-us/General-info/Types-of-control-reserve/Frequency-Containment-Reserve}{Frequency Containment Reserve by regelleistung.net, the official market portal for Germany}, accessed: 20.11.2023} Unlike other balancing markets, only the provision of capacity is reimbursed without a price for energy since positive and negative activations are expected to balance out on average.\footnote{\label{ftn:fcr_nextkraftwerke}\href{https://www.next-kraftwerke.de/wissen/primaerreserve-primaerregelleistung}{Definition Frequency Containment Reserve (FCR) by Nextkraftwerke}, accessed: 20.11.2023} The demand for the reserve is determined by a potential outage of the largest two power generators in the synchronous region and split across the participants.\footnote{ibid.} Activation quantities for similar products are described in the literature as negligible, like the Fast Frequency Response (FFR) and disturbance (FCR-D) products in the Nordic synchronous area \citep{thingvad_economic_2022}. \citet{saretta_electrolyzer_2023} state that FCR activations are not energy intensive, and activation payments can be negligible.

%Based on this, we create a case to make the power available to the FCR market for each of the six daily four-hour blocks. We model bids as a symmetrical product; hence, the battery reserves the offered power in both directions, upward and downward, since we do not know in which direction we might get activated beforehand. An appropriate SoC of the storage ensures that activations in both directions are feasible. We model no activations and assume balanced activation quantities at a low volume during each four-hour interval. Unused capacity can be used for spot market trading. We impose no limits on the allocation ratio of each market and let the model determine the optimal ratio. 

%Implementation
The problem is implemented in Julia, where we utilise the sddp.jl package by \citet{dowson_sddpjl_2021}. We approximate the optimal trading policy based on the mathematical problem formulation in Section \ref{sec:math_model} and the approximated Markov chain from Section \ref{sec:estimating_markov_chain}. From the data curation, we have 1081 individual price level combinations and 73 state variables, which implicitly define the Markov chain's approximately $4$ billion price paths. The SDDP algorithm takes about 45 minutes to estimate the policy for a subset of $3 \cdot 10^{6}$ combinations, using 18,000 iterations (forward and backward passes of the SDDP algorithm) with an Apple M1 Pro notebook processor. We found a CPU utilization of around 10\,\% in serial operation, indicating room for improvement by parallelising the training algorithm. Parallelisation requires a full model in working memory for each instance, which quickly reaches the hardware limits of the notebook. Moving the calculations to a server is possible and recommended for more advanced implementations. For the multi-market optimisation, we add a stopping criterion to the SDDP algorithm, which stops the algorithm if the upper bound improves with an absolute of less than 0.1 over the last 3000 iterations after an initial 5000 iterations. Otherwise, the algorithm runs until 18,000 iterations. The first trading day contains only the ID market actions and is omitted for a fair comparison of markets in the result tables.

To illustrate the model's functioning, we simulate the obtained policy in a simulation on fundamentals of the exemplary time period from 05.07.2022 to 08.07.2022. We tested multiple random periods and found consistent patterns with only minor variations in revenue and shares between markets that can be attributed to different arbitrage potentials for the prevailing prices. The 05.07.2022 - 08.07.2022 period sees relatively high volatility, which is favourable for spot-market participation.

\section{Results}
\label{sc:results}
In this section, we describe the trained policy from the SDDP algorithm and analyse the trading strategies across markets. 

%Our trading model consistently generates profits by using the battery in all markets and configurations. Therefore, we have successfully developed a trading strategy for battery storage that can operate optimally in high-volatile market environments. Hence, the method can be applied to evaluate the economics of batteries with multiple revenue streams and thus promote their adaptation in the market.

\subsection{Policy Evaluation}
\begin{figure}[htbp!]
\begin{minipage}{0.49\textwidth}
        \centering
        \includegraphics[width=\linewidth]{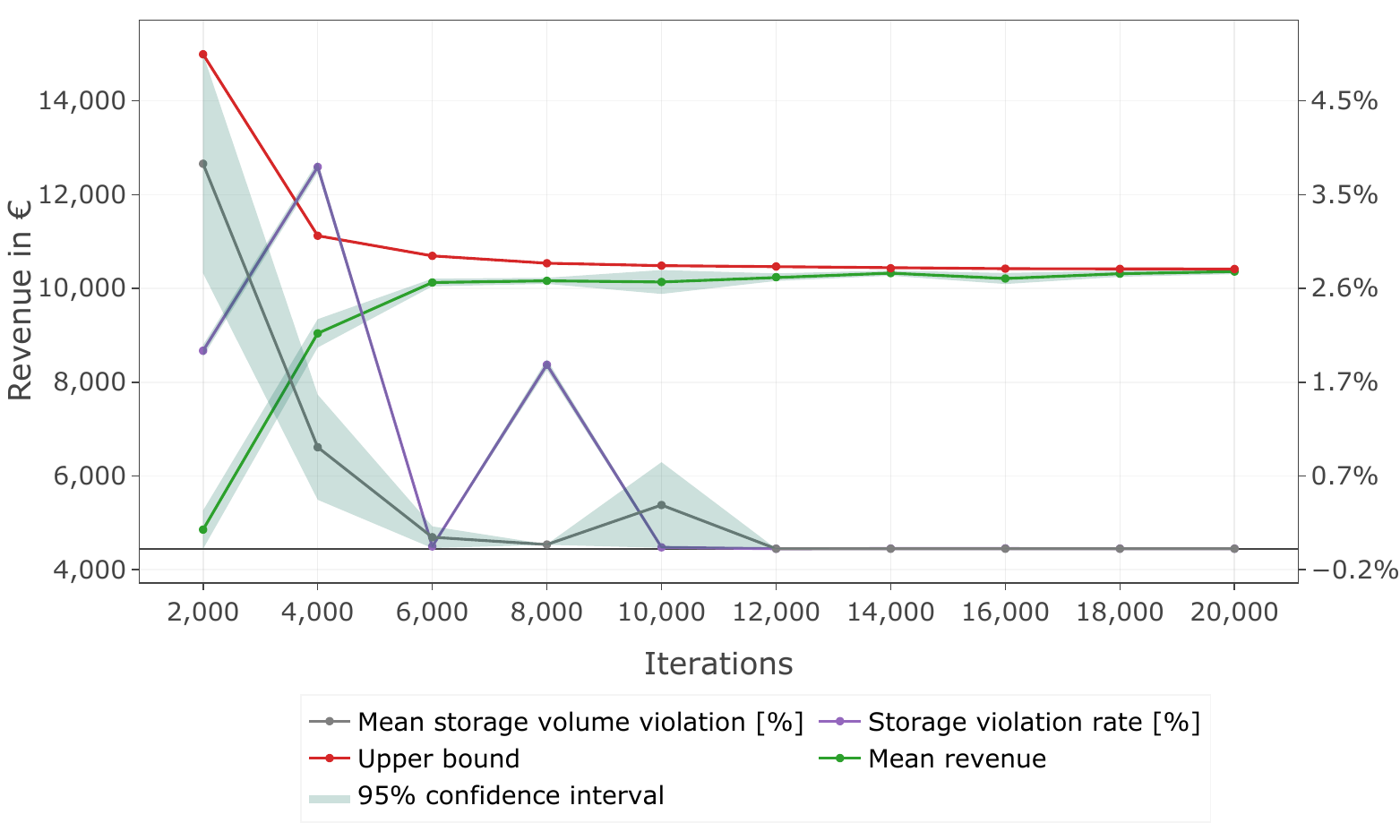}
        \caption{Objective values and storage violations at different iteration counts.}
        \label{fig:iteration_plot}
    \end{minipage}
    \hfill
    \begin{minipage}{0.49\textwidth}
        \centering
        \includegraphics[width=\textwidth]{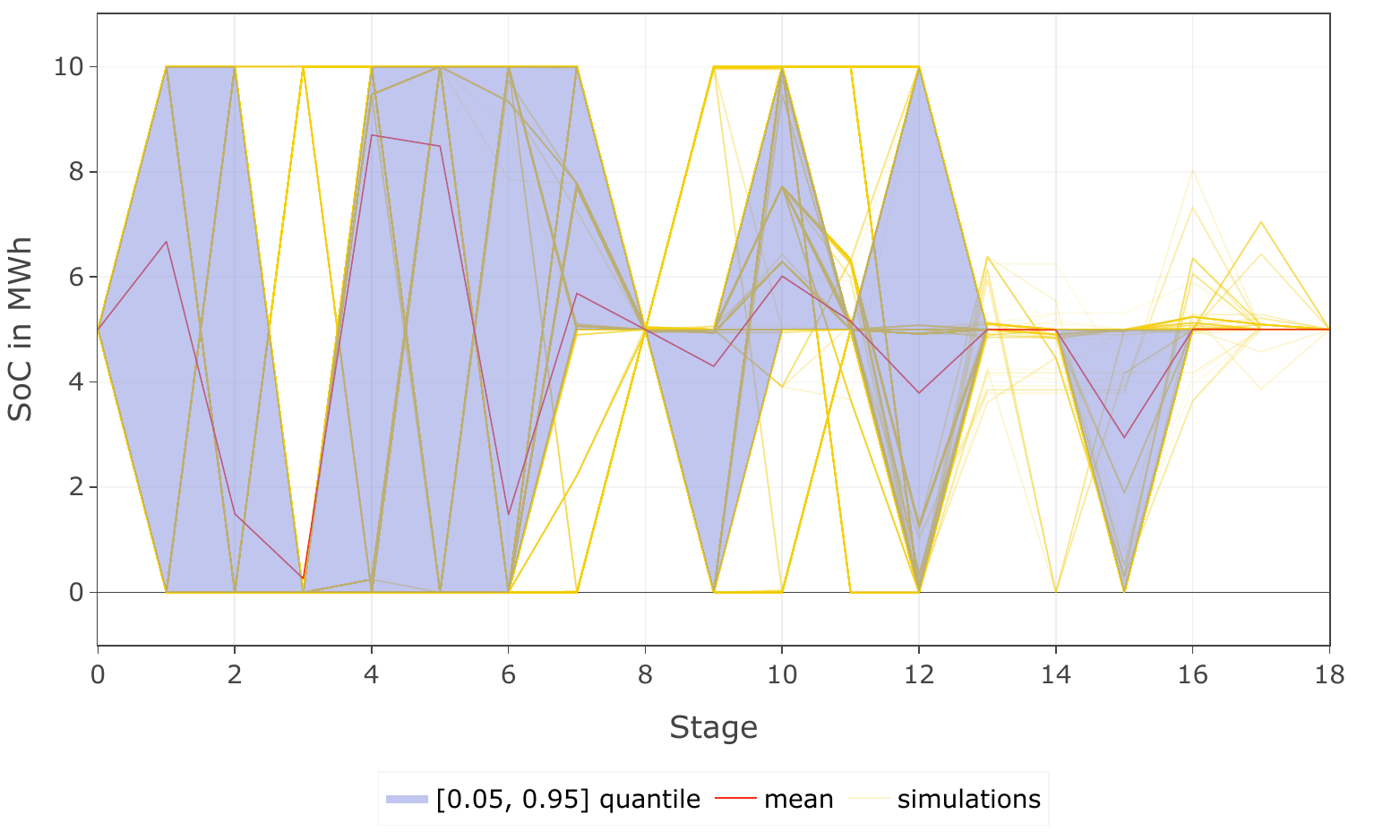}
        \caption{Battery SoC development at different stages with 18,000 iterations}
        \label{fig:SoC_over_stages}
    \end{minipage}
\end{figure}

We start by evaluating the computational performance of the most general policy, considering all markets. Figure \ref{fig:iteration_plot} shows the convergence of the SDDP algorithm as a function of the number of iterations.
We observe a policy improvement with higher iteration counts of training, resulting in a smaller gap between the true upper bound and simulations, as shown in Figure \ref{fig:iteration_plot}. We achieve a tight optimality gap after around 6,000 iterations. From 12,000 iterations on, no storage violations have been apparent, and only the trading strategy has improved. Further improvements flatten out with higher iteration counts while the marginal computational costs increase. Exemplary storage changes of a policy with 18,000 iterations with 10,000 simulation runs of simulated data are visualised in Figure \ref{fig:SoC_over_stages}. Based on the assessment of storage violations, the convergence of the true upper bound and the mean simulated revenue, we decided to continue with 18,000 iterations. The mean revenue of all simulations from three days of trading converges to an upper bound of 10,415\,\euro{}. The trading strategy thus results in profitable trading at calculation times that are within the market schedule for a trader in practice. In the following, we investigate how the trading strategy handles multi-market coordination.

\subsection{Multi-Market Coordination}

\begin{figure}[h!tbp]
    \centering
    \includegraphics[width=0.7\linewidth]{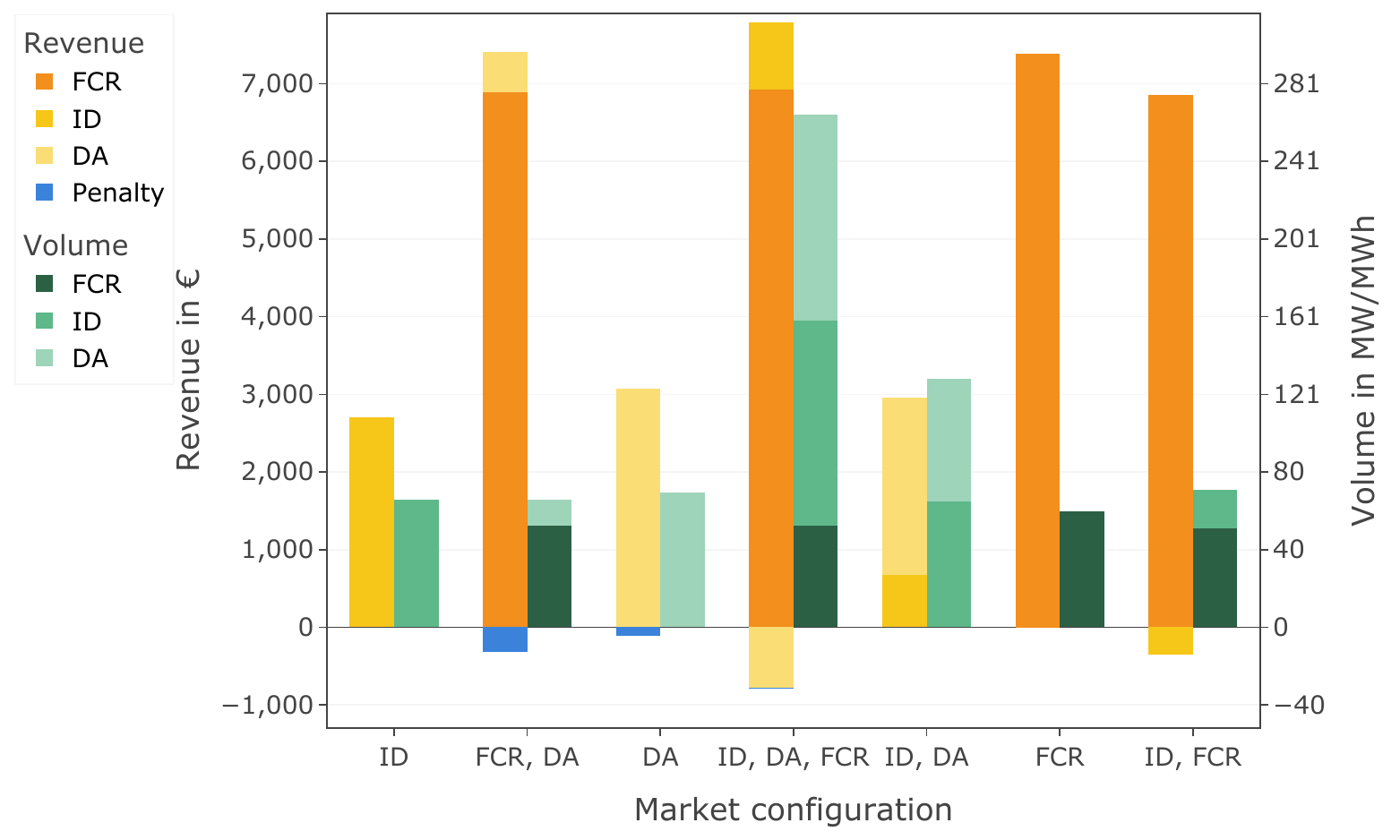}
    \caption{Mean revenue and market balance from two days on 10,000 in-sample simulations. The volume for FCR represents the reserved battery capacity in MW, whereas the volumes on DA and ID indicate traded volumes in MWh. These can contain opposite positions in DA and ID that do not go in physical delivery. Therefore, the configurations with both DA and ID markets show significantly larger traded volumes.}
    \label{fig:market_comparison_revenue_volume}
\end{figure}

To evaluate the benefits of participating in multiple markets, we develop a variety of policies that consider different market combinations.
% General observations and the reduction of risk by ID operation
Since only ID operations are possible on the first day (FCR and DA markets can only submit bid submissions for the next day), we compare the last two days of operation in the following. The revenues and balances of policies with different market participation are presented in Table \ref{tab:market_comparison_revenue_volume} and Figure \ref{fig:market_comparison_revenue_volume}. 
\begin{table}[htbp!]
\caption{Revenue and market balance from two days on 10,000 in-sample simulations with a 3,000\euro{}/MWh penalty for SoC and system-end-state violation}
\resizebox{\linewidth}{!}{\begin{tabular}{|ll|l|l|l|l|l|l|l|}
\toprule
\multicolumn{2}{|l|}{\textbf{Market configuration}} & \textbf{FCR} & \textbf{ID} & \textbf{DA} & \textbf{FCR, DA} & \textbf{ID, FCR}& \textbf{ID, DA}& \textbf{FCR, ID, DA} \\
\midrule
\textbf{DA}     & Revenue [\euro{}] &-      &-      &3092.4 &500.3  &-      &2066.0 &-780.9\\
                & Volume  [MW/MWh]  &-      &-      &69.8   &13.6   &-      &63.3   &106.8  \\
                & Balance  [MWh]    &-      &-      &0.0    &0.0    & -     &1.2    &-4.2   \\
\textbf{ID}     & Revenue           &-      &2705.9 &-      &-      &-351.4 &892.6  &863.9\\
                & Volume            &-      &65.9   &-      &-      &20.1   &65.3   &106.0  \\
                & Balance           &-      &0.0    &-      &-      &0.0    &-1.2   &4.2  \\
\textbf{FCR}    & Revenue           &7384.1 &-      &-      &6907.1 &6818.5 &-      &6928.2\\
                & Volume            &60.0   &-      &-      &52.8   &51.3   &-      &52.8 \\
\textbf{Penalty}& Cost [\euro{}]    &0.0    &0.0    &105.6  &238.9  &0.0    &0.0    &2.3  \\
\midrule
\textbf{Total}  & Revenue           &7384.1 &2705.9 &2986.8 &7168.5 &6467.1 &2958.6 &7008.8\\
                & Volume            &60.0   &65.9   &69.8   &66.4   &71.4   &128.6  &265.6  \\
\bottomrule
\end{tabular}
}\label{tab:market_comparison_revenue_volume}
\end{table}
The operational results show that a policy that exclusively bids in the FCR market outperforms all other policies in terms of revenue. Policies that add the DA market or DA and ID market result in similar revenues at higher volumes from additional spot market cycling but do not reach the same revenues as the exclusive FCR market operation.

% Combination of markets and inability to pick the corner solution
We observe that combinations of FCR with ID and DA markets neither see the exclusive allocation to the FCR market nor increased revenues over the exclusive FCR market case. The model avoids the corner solution of only participating in the FCR market, opting instead for a near-optimal solution. Nevertheless, they achieve results reasonably close to it while within a small optimality gap from increased computational complexity. Tests indicate that the optimality gap between pure FCR operations and combinations with spot markets decreases with higher iteration counts of the SDDP algorithm.

\begin{figure}[htbp]
    \centering
    \includegraphics[width=0.7\linewidth]{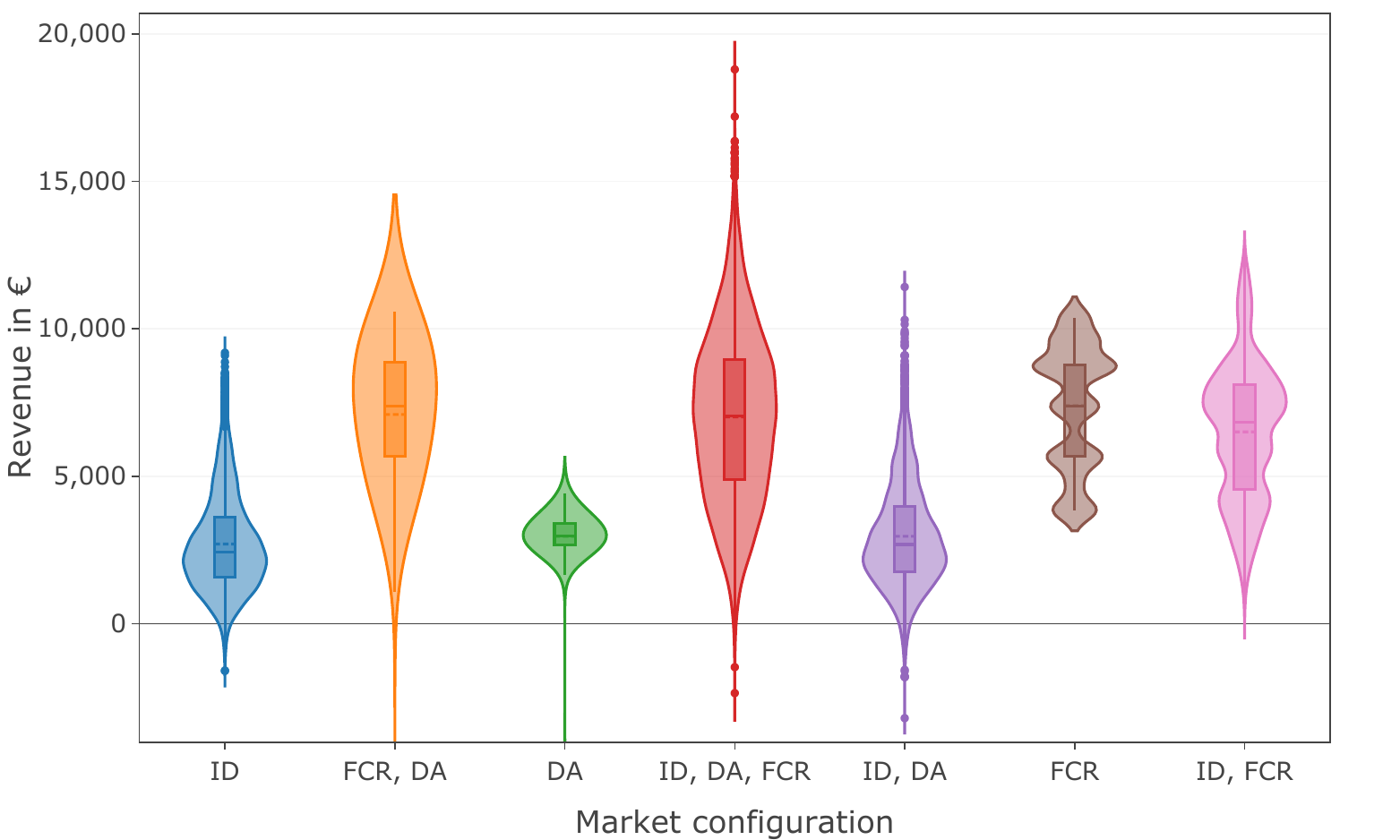}
    \caption{Revenue distribution of 10,000 simulations}
    \label{fig:violine_plot_revenue}
\end{figure}

The combination of ID and FCR markets marks an exception and leads to trading losses from ID participation, indicating non-beneficial coordination. Adding the ID market typically fails to provide additional value; a mix of ID and DA markets generates similar revenue to pure DA operation, while pure ID operation yields slightly lower results. Despite the ID market's higher volatility, its operation doesn't add revenue, likely due to unpredictability and limited cycling potential in our model's four-hour time resolution. However, ID operation and its combinations effectively reduce constraint violations and, thus, penalties.
% Further insights from the revenue distributions

Figure \ref{fig:violine_plot_revenue} shows the revenue distributions from the policy simulations. ID operation extends both tails of the revenue distribution, while pure DA operation results in a symmetrically bell-shaped distribution around the mean, marked by a long negative tail due to penalized storage violations. The participation in multiple markets can enhance tail revenues significantly compared to individual markets.

Trading volumes increase significantly when combining markets. We observe opposing trading patterns of ID and DA markets, indicating that the trading policy exploits arbitrage trading strategies. We assess the extent of arbitrage trading with two different metrics. The \textit{direction} metric is defined as:
\begin{align}
    \textit{direction} &= \frac{ \sum_{t \in \mathcal{T}} y^*_t}{\sum_{t \in \mathcal{T}} \left( y^{DA}_t+ y^{ID}_t\right)}, \label{eq:metrics_direction_def}  
\end{align}
where
\begin{align}
    y^*_t &= \begin{cases}  \min  \{|y^{DA}_t|, |y^{ID}_t|\} & \quad \text{if }  y^{DA}_t \cdot y^{ID}_t < 0,  \\
    0 & \quad \text{otherwise}, \end{cases} \quad  t \in \mathcal{T}.  
\end{align}
The intuition behind this metric is that the minimum of both trades with opposing signs is covered by the other market and doesn't affect the storage balance; therefore, arbitrages between markets. Moreover, the \textit{feasibility} metric calculates the storage-bound violations caused by DA and FCR trades that must be balanced in the ID:
\begin{align}
    \textit{feasibility} &= \frac{ \sum_{t \in \mathcal{T}} y^{**}_t}{\sum_{t \in \mathcal{T}} \left( y^{DA}_t+ y^{ID}_t\right)}, \label{eq:metrics_feasibility_def}  
\end{align}
where
{\small
\begin{align}
     y^{**}_t &= \begin{cases} y^{DA}_t+y^{FCR}_t -SoC_t & \quad y^{DA}_t+y^{FCR}_t -SoC_t >0, \\
    |L + y^{DA}_t - y^{FCR}_t-SoC_t| & \quad L + y^{DA}_t - y^{FCR}_t-SoC_t <0,\\
    0 & \quad \text{otherwise},
    \end{cases} \quad  t \in \mathcal{T}.  
    \label{eq:metrics_def_2}
\end{align}
}
Our results indicate that, on average, 39.6\,\% of the combined market volume corresponds to arbitrage trading using the \textit{direction} metric, and 39.5\,\% using the \textit{feasibility} metric. There is a significant shift of volume from the DA to the ID market, resulting in a 4.2\,MWh surplus in the ID market over the three days.

%Based on the input data and the aggregation of four-hour time slices, we do not observe additional revenue from coordinated bidding. That is why we include the case with reduced FCR prices in the analysis to show the proof-of-concept for the SDDP strategies in the following case. 

\subsection{Modified FCR prices}
We found that trading solely on the FCR market yielded the best results with the given input data and a four-hour time resolution. This is surprising, as we expected a multi-market strategy to be more profitable. One reason for this could be the model's uniform time resolution, which contrasts with real-world markets where the DA and ID markets operate on finer hourly and quarter-hour resolutions. This difference allows traders to capture more volatility and price spreads, potentially enhancing profitability.

\begin{figure}[htbp]
    \centering
    \includegraphics[width=0.7\linewidth]{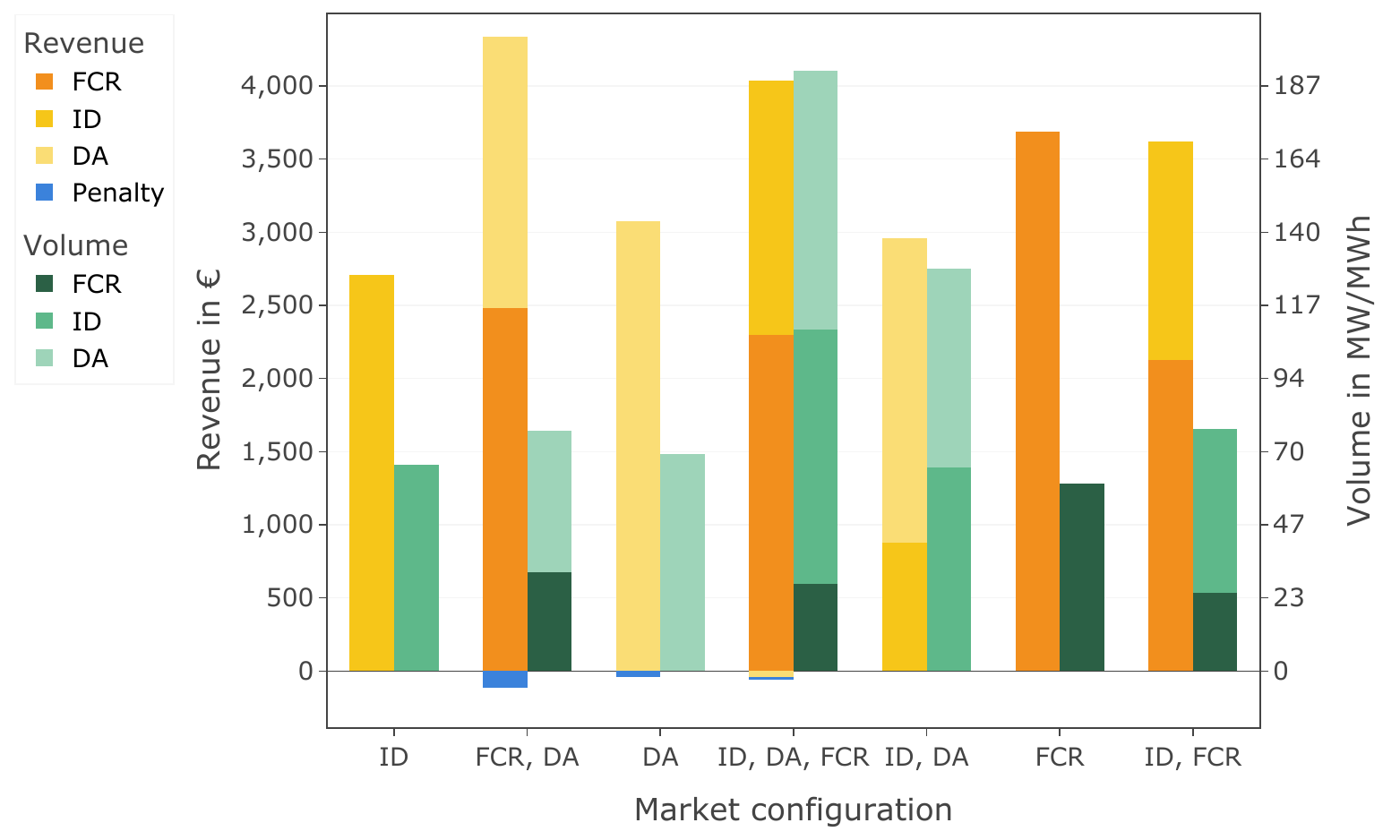}
    \caption{Mean revenue and market balance from two days on 10,000 in-sample simulations on 50\,\% decreased FCR prices}
    \label{fig:market_comparison_revenue_volume_low_FCR}
\end{figure}

To address this fundamental limitation of time resolution and investigate the ability of our bidding model to approximate a trading policy with market coordination, we create an additional analysis with reduced FCR price levels by 50\,\% to compensate for reduced cycling on the spot markets, allowing us a better understanding of the intrinsic market dynamics and the benefits of coordinated bidding. The results are shown in Table \ref{tab:market_comparison_revenue_volume_low_FCR_price} and Figure \ref{fig:market_comparison_revenue_volume_low_FCR}. 
\begin{table}[htbp]
\caption{Mean revenue and market balance from two days trading at a 50\,\% reduced FCR prices on 10,000 in-sample simulations with a 3,000\euro{}/MWh penalty for SoC and system-end-state violation}
\resizebox{\linewidth}{!}{\begin{tabular}{|ll|l|l|l|l|l|l|l|}
\toprule
\multicolumn{2}{|l|}{\textbf{Market configuration}} & \textbf{FCR} & \textbf{ID} & \textbf{DA} & \textbf{FCR, DA} & \textbf{ID, FCR}& \textbf{ID, DA}& \textbf{FCR, ID, DA} \\
\midrule
\textbf{DA}     & Revenue [\euro{}] &-      &-      &3073.4&1852.1&-      &2079.0&-40.6\\
                & Volume  [MW/MWh]    &-      &-      &69.5  &45.1  &-      &63.4  &82.7  \\
                & Balance  [MWh]    &-      &-      &0.0    &0.0   & -     &1.2   &-3.1   \\
\textbf{ID}     & Revenue  &-      &2705.9&-      &-      &1493.3 &879.9 &1740.8\\
                & Volume            &-      &65.9  &-      &-      &52.6  &65.3  &81.4  \\
                & Balance           &-      &0.0    &-      &-      &0.0    &-1.2  &3.1  \\
\textbf{FCR}    & Revenue  &3689.9&-      &-      &2484.2&2124.4&-      &2298.5\\
                & Volume            &60.0   &-      &-      &31.7  &25.0  &-      &28.0 \\
\textbf{Penalty}& Cost [\euro{}]    &0.0    &0.0    &43.8  &115.6 &0.0    &0.0    &18.5  \\
\midrule
\textbf{Total}  & Revenue           &3689.9&2705.9&3029.6&4220.7 &3617.7&2958.9 &3980.3\\
                & Volume            &60.0   &65.9  &69.5  &76.8   &77.6  &128.7 &192.1  \\
\bottomrule
\end{tabular}
}\label{tab:market_comparison_revenue_volume_low_FCR_price}
\end{table}

\begin{figure}[htbp]
    \centering
    \includegraphics[width=0.7\linewidth]{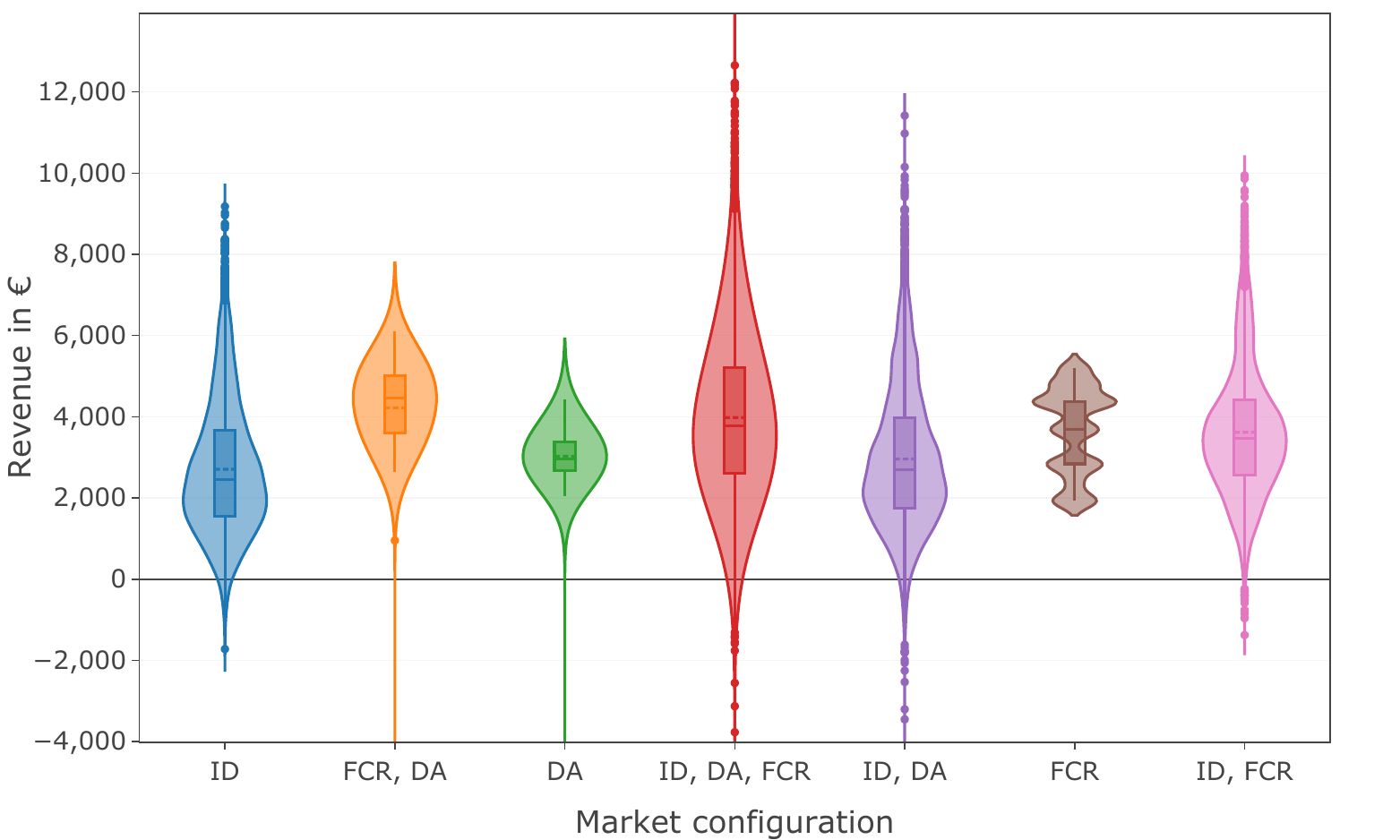}
    \caption{Revenue distribution of 10,000 simulations on 50\% decreased FCR prices}
    \label{fig:violine_plot_revenue_low_FCR}
\end{figure}
We see that the profitability ranking of the individual markets remained intact, with ID, DA, and FCR markets in ascending order. The revenues and volumes of arrangements containing only DA and ID markets see little variations from the price changes and stay at the original cases' levels within the sampling variation. Policies that include the FCR market have reduced revenues caused by the price adjustments. Most strikingly, we now observe multi-market trading policies outperforming individual-market policies by increasing revenue through higher allocations to spot markets and lower allocations to the FCR market. Combining DA, ID, and FCR markets improves revenue, with the FCR and DA combination yielding the highest revenue at an additional 530.8\euro{}, or 12.5\,\%, compared to just FCR market. Combining all markets increases revenue by 5.7\,\% over the best individual market option. Moreover, storage violation penalties decrease compared to operating solely in the DA market due to a shift in volume from DA to ID. The arbitrage volume between ID and DA declines significantly.
\section{Discussion}
\label{sc:discussion}
%This section critically reflects on our findings, placing them within the context of existing literature. We discuss the implications of our assumptions and configuration choices and highlight directions for future research across five categories.

\paragraph{Time resolution and battery properties} The results demonstrate that our chosen methodology is appropriate for adapting to the three-market environment, effectively capturing the stochasticity of the markets, even during periods of high volatility. The solution method can handle complexity within a time frame that is suitable for practitioners. However, the case of modified FCR prices highlights a key dilemma in modelling battery storage revenues: the model resolution significantly affects performance. A four-hour resolution underestimates revenue potential in wholesale markets, as spot market revenue relies on time arbitrage. Limiting cycling through a low resolution restricts spot market profits, while reserving capacity in the balancing market is cycling independent and thus yields higher returns in the investigated case study. We further use average prices in the four-hour intervals, which smoothened price spikes, especially in the ID market. The battery's power rating is high enough to complete a full cycle within an hour. An operator executing fast cycling would likely focus more on the technical properties of the battery, like degradation, temperature and efficiency losses caused by heavy cycling in the modelling. An even faster cycling in the 15-minute ID market with periodic (and storage level dependent) fast charging could further improve revenues but comes at the expense of high computational costs and additional (technical) constraints. Additionally, the battery's power/capacity configuration (10MW/10MWh) can influence the revenues of the different markets. Larger power ratings favour flexibility provision in capacity markets, i.e. FCR. A battery in a 0.5C configuration (10MW/20MWh) would make the same FCR revenues but considerably higher revenues on the spot markets. The power and capacity ratio can thus be investigated as a sensitivity parameter in future work. 

\paragraph{Value of coordination} In line with \citet{kongelf_portfolio_2019}, we do not find significant coordination values in expectation at the chosen modelling resolution. However, the revenue distribution shows spikes in volatility, and significant volume shifts between DA and ID markets suggest arbitrage opportunities. With lower FCR prices that reduce individual-market dominance, we observe an additional value in coordinating spot and balancing market bids, as these prices bring capacity reservation revenues and spot market time arbitrage closer together, partially offsetting reduced cycling from the model's lower temporal resolution.

Despite the dominance of one individual market, this are no signs of unwanted incentives as in \citet{boomsma_bidding_2014} since allocating all capacity in the FCR market is limited by the FCR size in Germany. The market size is currently only 600MW and will soon be saturated as a revenue stream with an increase in large battery projects, whereas DA size and ID size are not expected to be saturated in the near future. That means that cannibalisation effects with decreasing prices will leave no alternative to splitting up revenues that share the battery capacity and stack multiple revenue streams. Contrary to \citet{lohndorf_value_2022} but in line with \citet{heredia_optimal_2018}, we find no increases in profits from trading the higher volatility of the ID market but notice positive effects on imbalances. However, no common benchmark exists in the academic sphere, and results depend on the case study and input data.

\paragraph{Solution method} A limitation of SDDP in large-scale optimisation is its sampling of stochastic scenarios from the Markov chain, which can lead to some price paths being overlooked and result in near-optimal solutions. Our example illustrates this: while full participation in the FCR market is best in expectation, the model does not fully converge to this solution but instead identifies near-optimal market combinations. Increasing the number of iterations narrowed the gap to the optimal solution, but single FCR participation was never fully reached. Given the high complexity of the problem and the curse of dimensionality, we navigate the solution space by SDDP and evaluate at most 5,107,075 of the Markovian price paths ($0.000124\,\%$ of all combinations). The achieved optimality gap of 5\,\% compared to the exclusive FCR market solution is deemed acceptable for the considered academic investigation. We are aware that in the current state, with the four-hour resolution and given the parameters of the case study, no significant improvement compared to the single market strategy is realized. Slack variables used to stabilize the SDDP algorithm may negatively impact results, as some simulated policies led to substantial negative revenues due to capacity violations. This reveals a conflict between adopting policies that ensure feasible actions and maintaining a realistic cost framework. Our fixed imbalance penalty could be overly conservative compared to short-notice ID settlements and potential balancing costs from the grid operator, risking premature cutoff of trades when storage constraints are violated.

\paragraph{Capacity activations} Limitations apply to imbalances in the battery's storage energy due to capacity activations in the FCR market. Although these imbalances occur with low probability and involve minimal energy, they can disrupt the battery's energy balance, potentially triggering costly short-notice buy actions on the market. It is interesting to see how these additional costs rebalance the allocation of volumes between the markets against the high expected revenues. To the best of our knowledge, no publicly available data on FCR activations exists, rendering it difficult to perform a critical validation. 

\paragraph{Storage as a price taker} The price-taker assumption in markets with limited liquidity needs careful consideration. While the DA market is highly liquid, the ID and FCR markets may have price influences, resulting in lower volatility and reduced revenues\footnote{Non-peer-reviewed market analyses for the DA in grey literature question the price taker assumption. See for example: \href{https://www.regelleistung-online.de/preiseffekte-durch-den-ausbau-von-batteriespeichern-teil-3-arbitrage-in-der-day-ahead-auktion/}{regelleistung-online: Preiseffekte durch den Ausbau von Batteriespeichern – Teil 3: Arbitrage in der Day-Ahead Auktion}, accessed \DTMdate{2023-07-08}. \nopagebreak}. Considering price impacts would require the introduction of binary variables. A solution framework, such as Stochastic Dual Dynamic Integer Programming, could handle this but comes at significantly higher computational costs. We note that the battery size in the case study is not large enough to expect a significant price impact from its market participation, given the liquidity of the markets in Germany. To analyse larger storage assets and their price impact, we refer to \citet{barbry_robust_2019}. 

\paragraph{Scenario generation} More advanced data preparation and scenario reduction techniques could enhance revenues when constructing the Markov chain. Employing k-means clustering may overlook complex patterns within the underlying data. Although we tried to capture these time-dependent dependencies with our econometric models and time variables, limitations remain in some cases. Since these variations are included in the stochastic component, further pattern recognition with advanced clustering algorithms, like Density-based spatial clustering of applications with noise (DBSCAN), could be a direction for future research.

\section{Conclusion}
\label{sc:conclusion}
We contribute to the literature on coordinated trading by battery storage in electricity markets. Our study includes the DA and ID spot markets, along with the FCR balancing market, which has received limited attention but is relevant in practice. We present a case study on coordinating a 10MW/10MWh battery storage in Germany across three markets. To explore multi-market trading, we developed a stochastic multi-market bidding model and solved it using SDDP within reasonable computational times while adhering to time coupling and storage boundary constraints. Additionally, we developed and calibrated econometric price models for the FCR, DA, and ID markets, incorporating a simple yet effective separation of market fundamentals and stochasticity for practical implementations. Our findings reveal that introducing quantiles of residual demands as a regressor for scarcity effects significantly improved our econometric models, particularly in the FCR market. Our stochastic bidding model proves to be well-suited for the problem at hand. It consistently generates expected profits across all markets and configurations, can be (re)trained between market scheduling steps and leverages synergies from multi-market coordination. However, when considering the FCR balancing market in isolation, it dominated expected revenues compared to combinations of wholesale markets. This suggests that a profit-maximizing operator during the German energy crisis would likely prefer this corner solution over market coordination, although the four-hour temporal resolution of our model impacts spot market trading. To compensate for this limitation, we decreased the FCR price level and observed a value of coordination between spot and balancing markets of up to 12.5\,\%. Our methodology can be used to provide valuable insights into the economic feasibility of energy storage deployment within the German energy sector, offering a forward-looking perspective on the role of storage technologies in the evolving energy landscape. We contribute to the literature by discussing model resolution and storage dimensioning and recommend that future research on multi-market coordination with battery storage, with a focus on higher-resolution models with shorter time intervals, the interaction between offered volumes and prices and the extension of coordination to additional products. In practice, coordination by battery storage might further motivate investigations into incentive compatibility of the current market setup, particularly since profitability equilibria across the market segments compete for the flexibility of dispatchable assets.

\section*{Acknowledgements}
The Research Council of Norway funded this work via project number 320789. We acknowledge funding from the KIT House of Young Scientists (KHYS) by a travel budget for visiting KIT in Karlsruhe. Special thanks to Benedikt Krieger for his support in data compilation and Oscar Dowson for the continuous support of the SDDP.jl package. We thank colleagues and participants of the NTNU's Energy System Seminar and the Transatlantic Infraday Conference 2023 in Paris for their input and valuable feedback. 

\section*{Declaration of competing interest}
The authors declare that they have no known competing financial interests or personal relationships that could have appeared to influence the work reported in this paper at the time of writing.

\section*{Declaration of generative AI and AI-assisted technologies in the writing process}
During the preparation of this work, the authors used Grammarly in order to
improve orthography and readability. After using this service, the authors reviewed
and edited the content as needed and take full responsibility for the content of the published
article.

\section*{CRediT}
\noindent\textbf{Paul E. Seifert:} Conceptualization, Data curation, Formal analysis, Investigation, Methodology, Software, Validation, Visualization, Writing – original draft, Writing – review \& editing.
\textbf{Emil Kraft:} Conceptualization, Data curation, Formal analysis, Investigation, Methodology, Supervision, Visualization, Writing – review \& editing.
\textbf{Steffen J. Bakker:} Conceptualization, Investigation, Methodology, Supervision, Writing – review \& editing.
\textbf{Stein-Erik Fleten:} Conceptualization, Investigation, Methodology, Supervision, Writing – review \& editing.

\bibliographystyle{model5-names}
%\bibliographystyle{elsarticle-num}

\bibliography{references.bib}

\newpage
\appendix
\section{Appendix}
\subsection{Multi-variate k-means clustering of prices}\label{sec:clustering}
\begin{figure}[h!tbp]
    \begin{minipage}{0.49\textwidth}
        \centering
        \includegraphics[width=\textwidth]{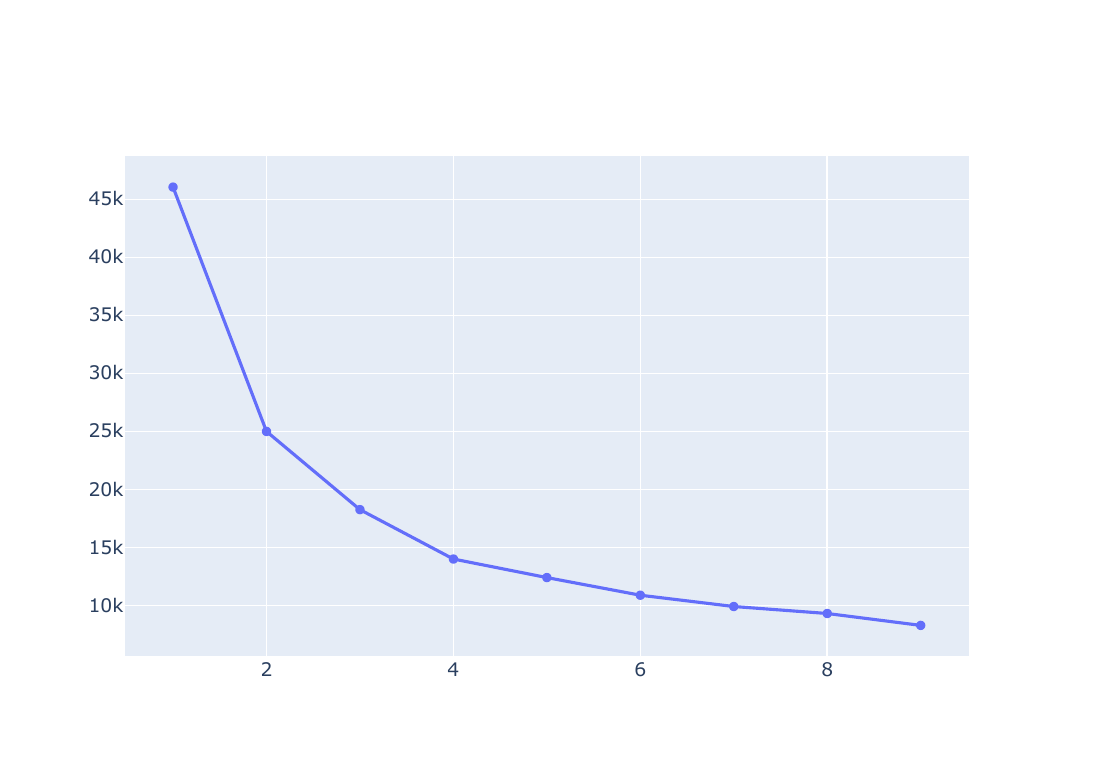}
        \caption{Elbow-plot of the multi-variate k-means clustering of DA prices}
        \label{fig:DA_elbow}
    \end{minipage}
        \hfill
    \begin{minipage}{0.49\textwidth}
        \centering
        \includegraphics[width=\textwidth]{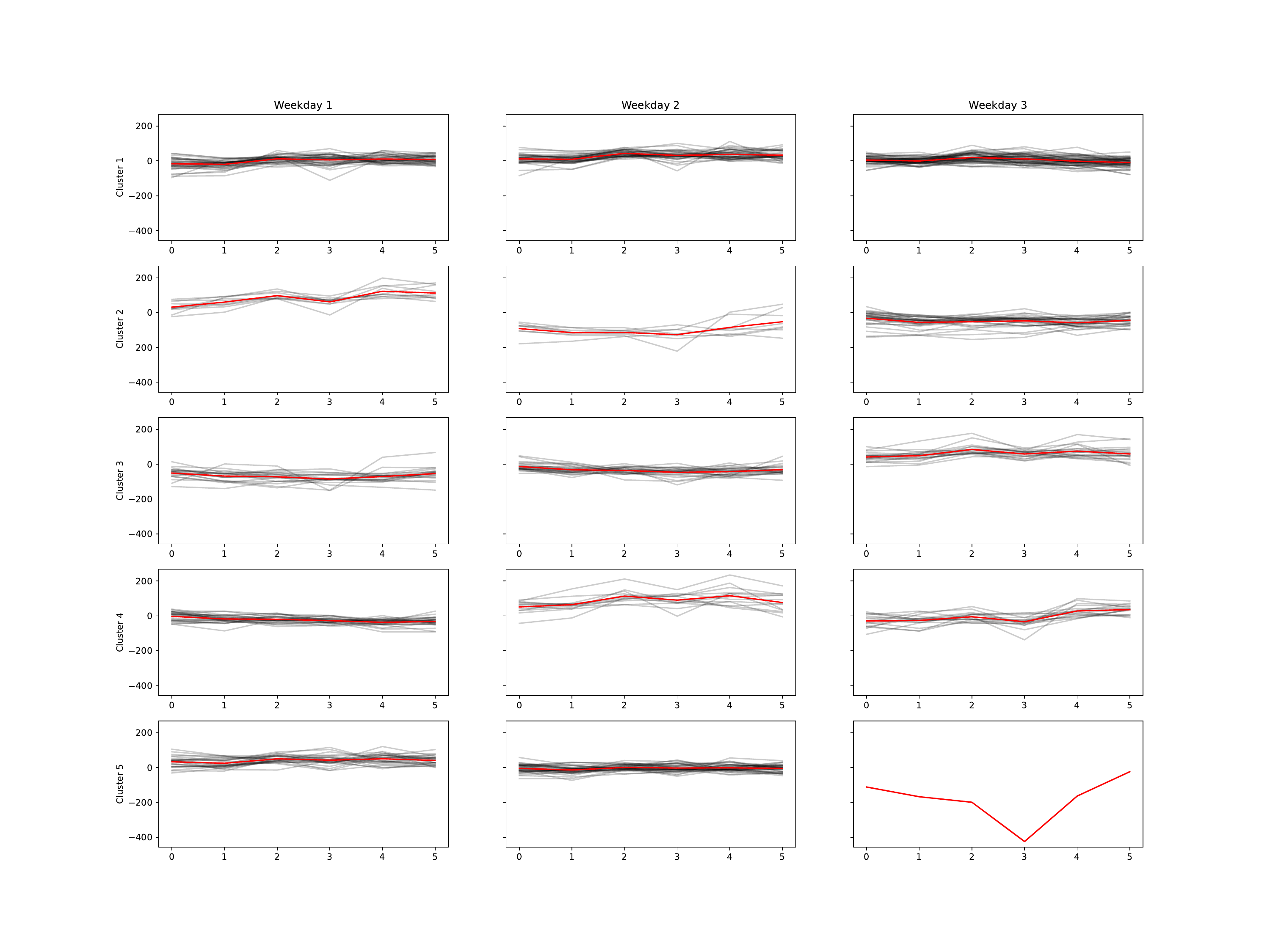}
        \caption{Results of k-means clustering of stochastic DA price components}
        \label{fig:DA_cluster}
    \end{minipage}
\end{figure}

\begin{figure}[h!tbp]
    \begin{minipage}{0.49\textwidth}
        \centering
        \includegraphics[width=\textwidth]{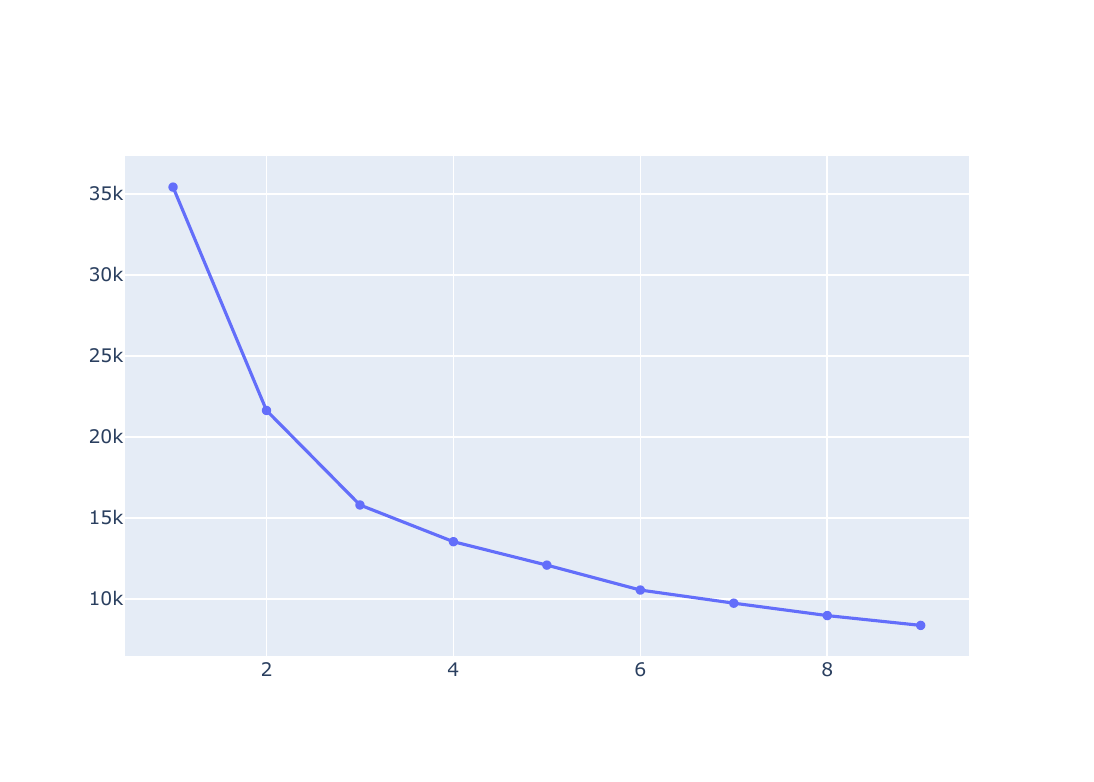}
        \caption{Elbow-plot of the multi-variate k-means clustering of FCR prices}
        \label{fig:FCR_elbow}
    \end{minipage}
        \hfill
    \begin{minipage}{0.49\textwidth}
        \centering
        \includegraphics[width=\textwidth]{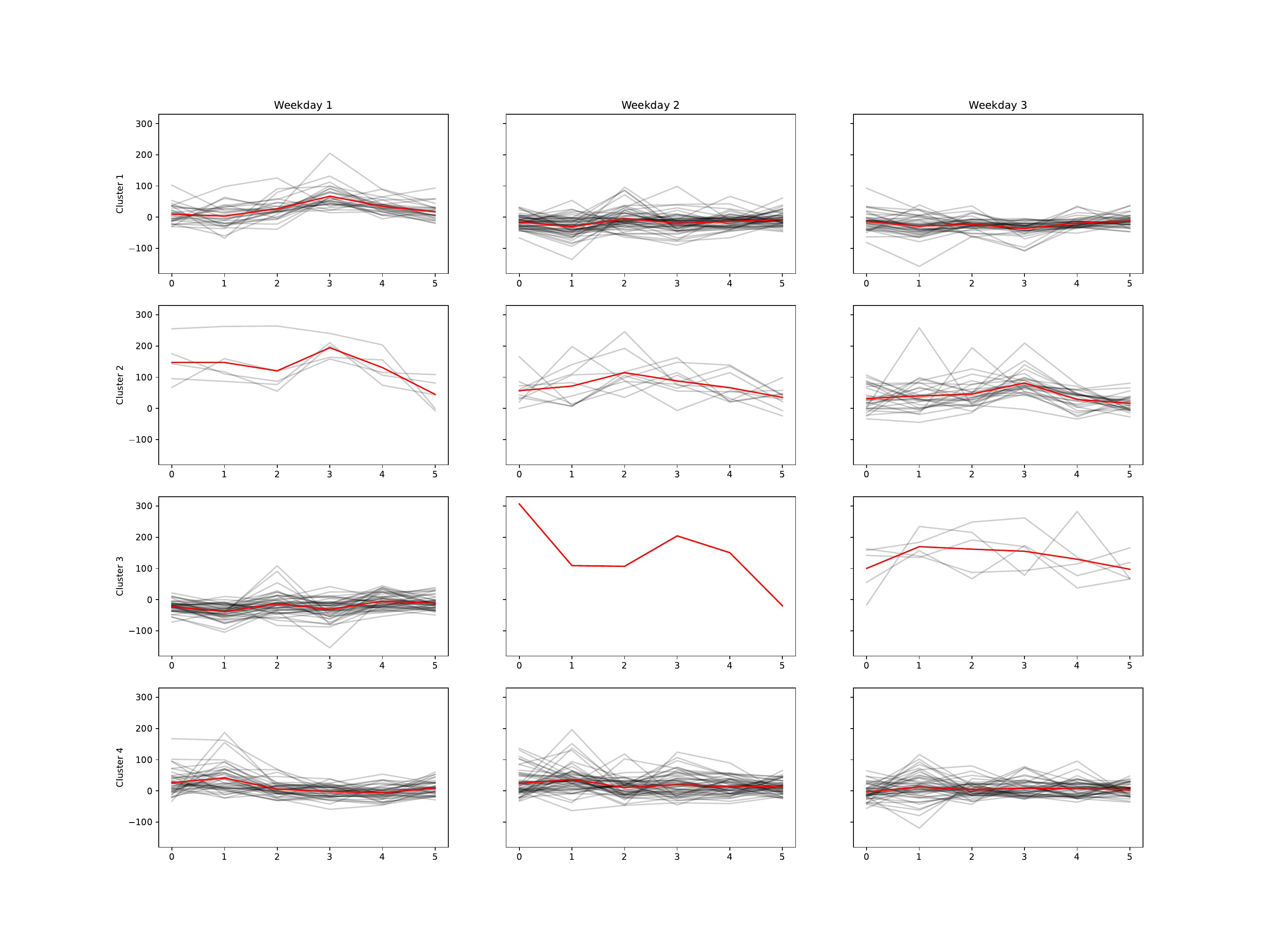}
        \caption{Results of k-means clustering of stochastic FCR price components}
        \label{fig:FCR_cluster}
    \end{minipage}
\end{figure}
\subsection{Market prices in the investigated period}
\begin{figure}[h!tbp]
    \begin{minipage}{0.49\textwidth}
        \centering
    \includegraphics[width=\textwidth]{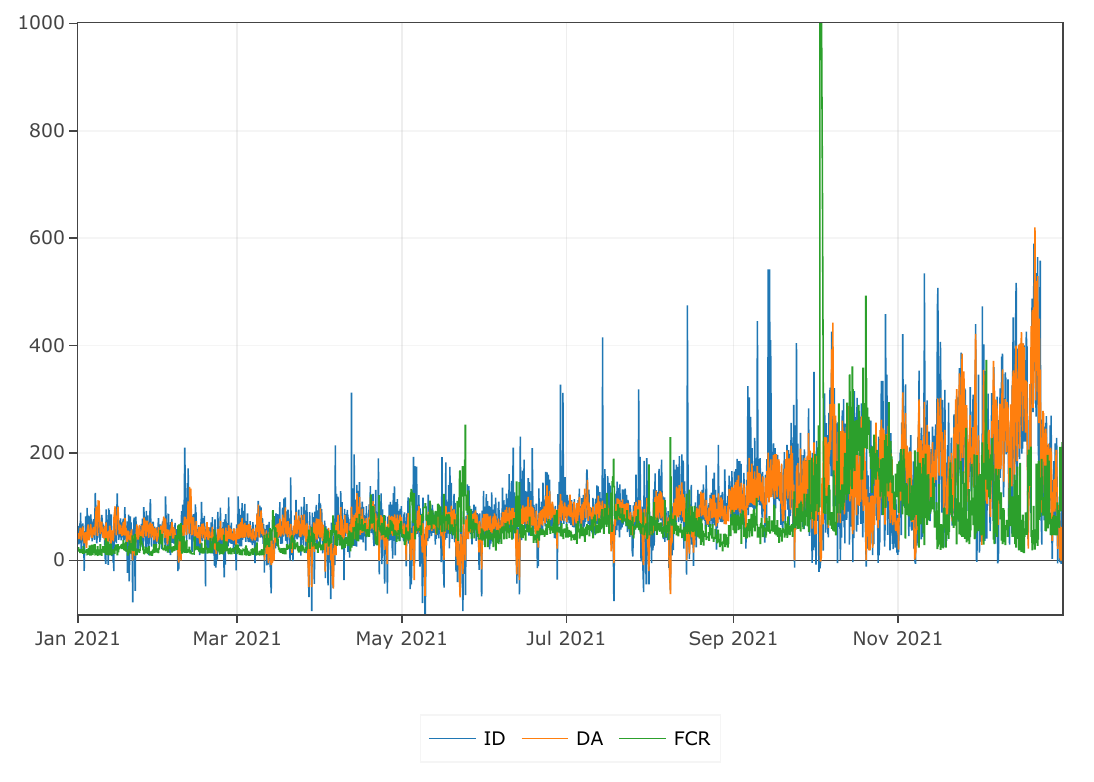}
    \caption{Prices of the DA, ID and FCR market of the year 2021 in Germany}
    \label{fig:prices_2021}
    \end{minipage}
    \hfill
    \begin{minipage}{0.49\textwidth}
        \centering
    \includegraphics[width=\textwidth]{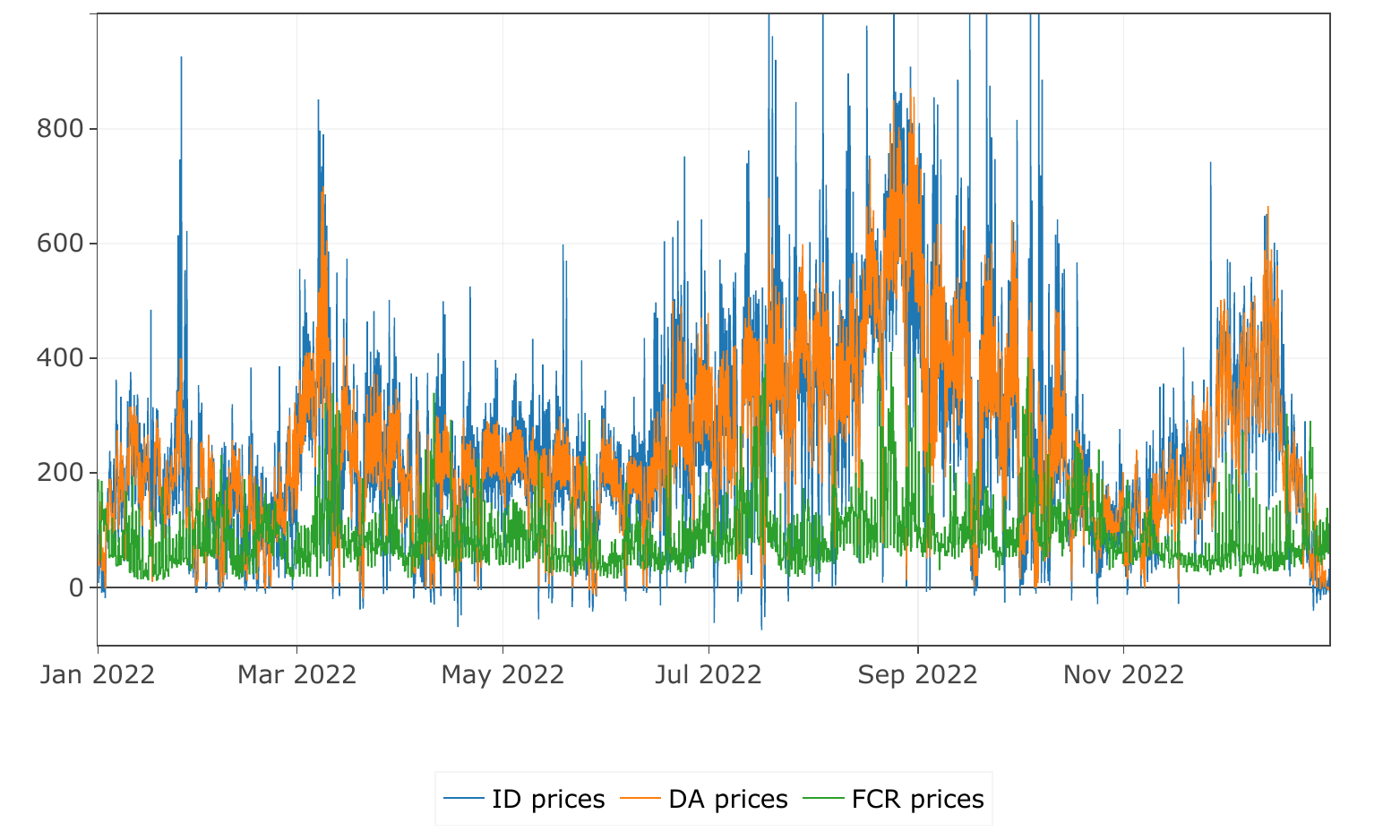}
    \caption{Prices of the DA, ID and FCR market of the year 2022 in Germany}
    \label{fig:prices_2022}
    \end{minipage}
\end{figure}

Market prices for the three markets of DA, ID and FCR in Germany are listed below for the years 2021 (Figure \ref{fig:prices_2021}) and 2022 (Figure \ref{fig:prices_2022}). They show a strong increase in the price level from the second half of 2021 and a general increase in volatility throughout 2022. 

\subsection{Selection of the Penalty Term}

\begin{figure}[htbp!]
    \begin{minipage}{0.49\textwidth}
        \centering
    \includegraphics[width=\textwidth]{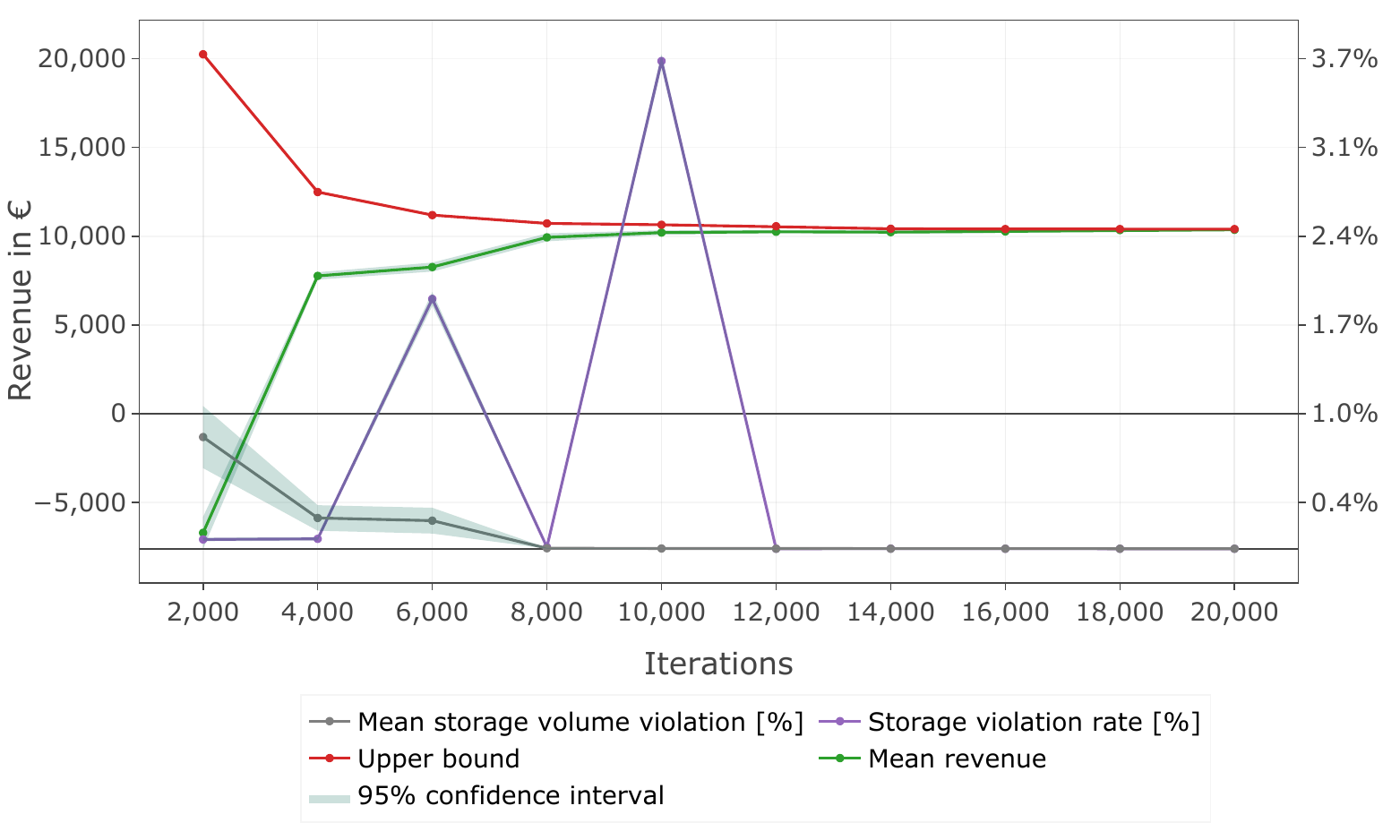}
    \caption{Objective values and storage violations at different iteration counts at a penalty of 10,000\euro{}/MWh}
    \label{fig:different_iterations_10000}
    \end{minipage}
    \hfill
    \begin{minipage}{0.49\textwidth}
        \centering
    \includegraphics[width=\textwidth]{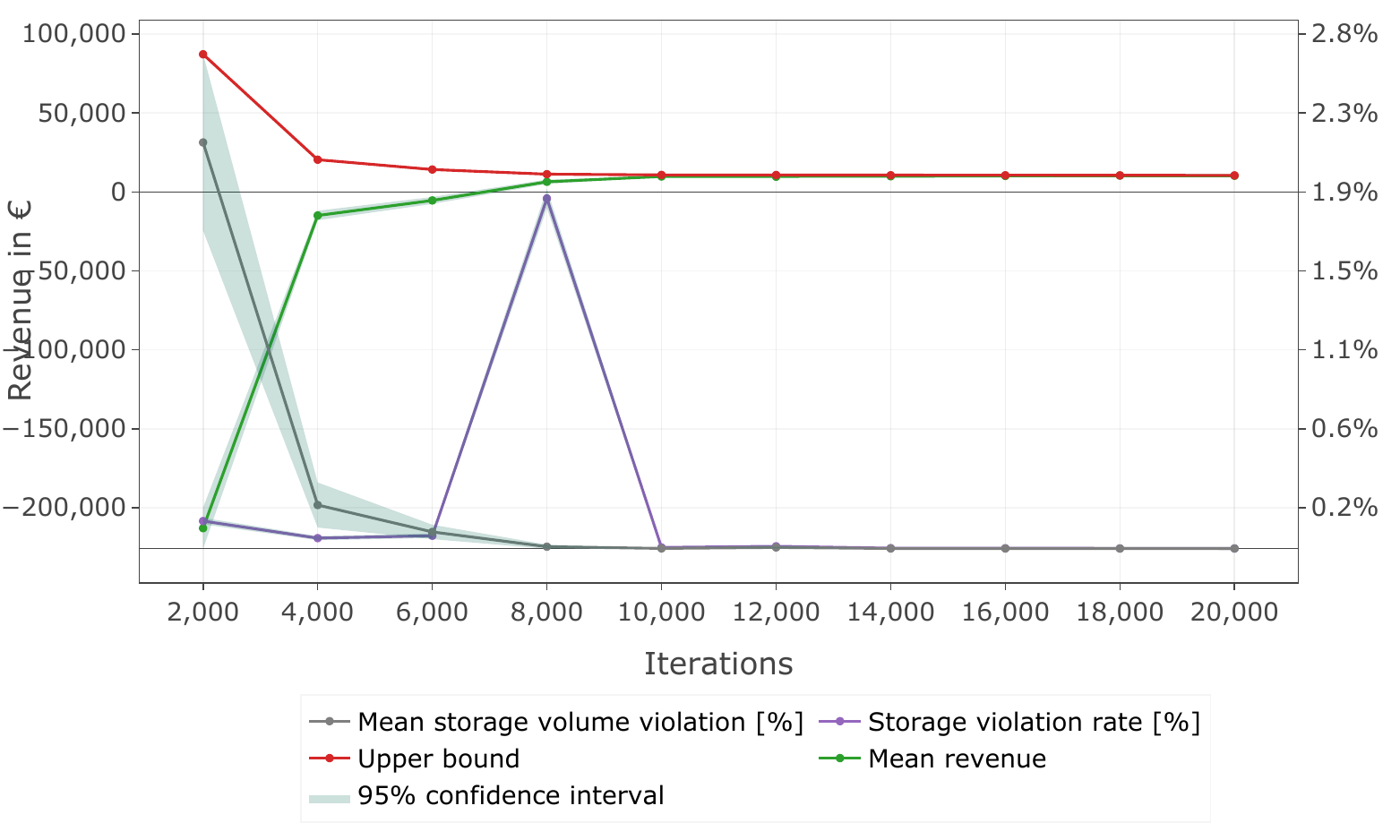}
    \caption{Objective values and storage violations at different iteration counts at a penalty of 100,000\euro{}/MWh}
    \label{fig:different_iterations_100000}
    \end{minipage}
\end{figure}

We have tested different configurations of storage violation penalties, including 100,000 \euro{}/MWh (Figure \ref{fig:different_iterations_10000}), 10,000 \euro{}/MWh (Figure \ref{fig:different_iterations_10000}), and 3,000 \euro{}/MWh (Figure \ref{fig:iteration_plot}), along with varying iteration counts. All configurations increase the storage violation penalty term and increase losses within simulations, especially at fewer iterations. After about 8,000 iterations, a violation magnitude of about 1\,\% for each violation can be observed. However, these minor violations occur more often. Notably, less-trained policies result in higher negative revenues due to storage limit violations, and we observe reduced convergence. The resulting policy at high iteration counts is comparable with lower penalty terms; therefore, we conclude that our lower penalty term of 3,000 \euro{}/MWh is sufficient and continue the rest of the investigation with it.
\label{ch:high_penalties}

\subsection{Removing the ID Constraint}
\label{ch:without_ID_const}
\begin{figure}[htbp!]
\begin{minipage}{0.49\textwidth}
        \centering
        \includegraphics[width=\linewidth]{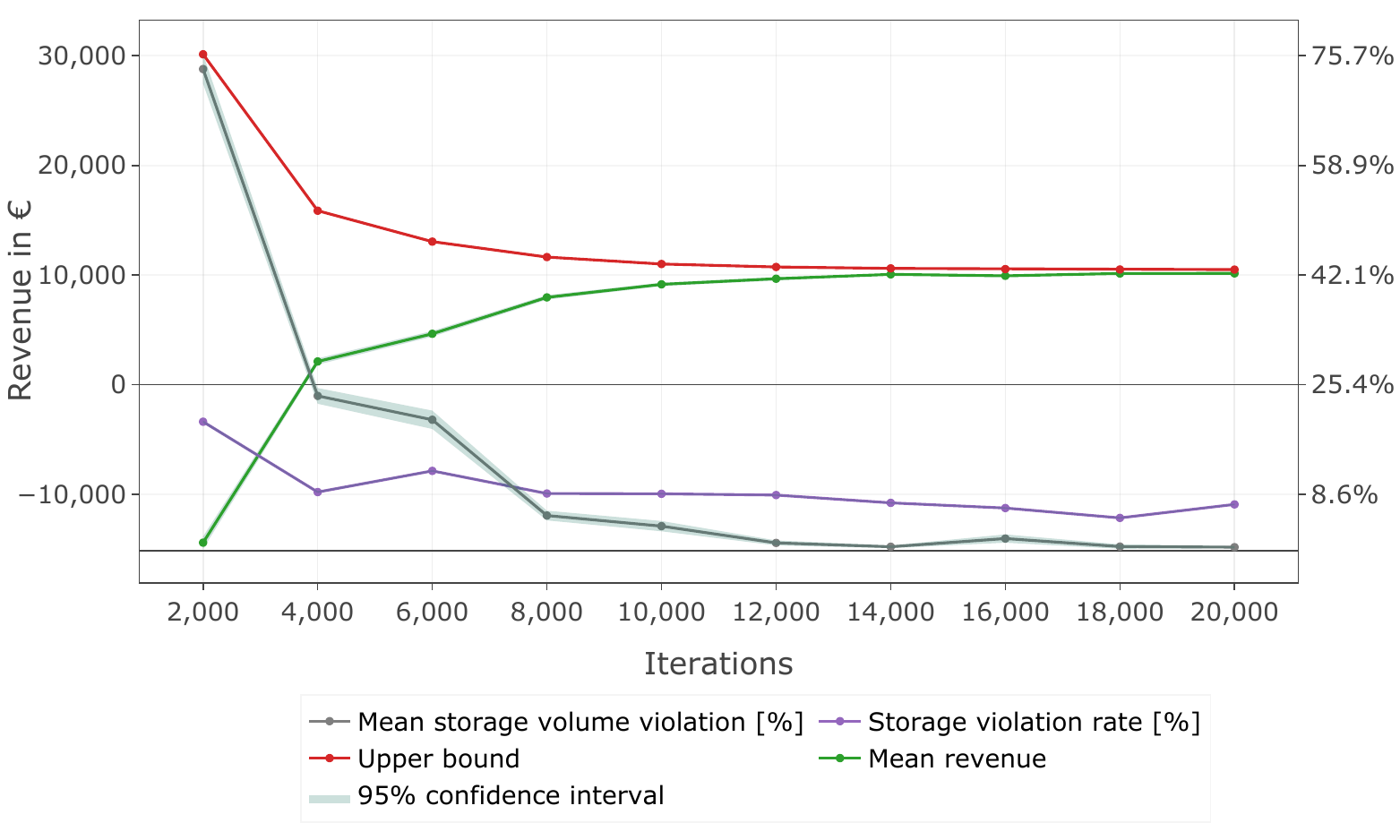}
        \caption{Objective values and storage violations at different iteration counts in the absence of constraints.}
        \label{fig:iteration_plot_without_ID_const}
    \end{minipage}
    \hfill
    \begin{minipage}{0.49\textwidth}
        \centering
        \includegraphics[width=\textwidth]{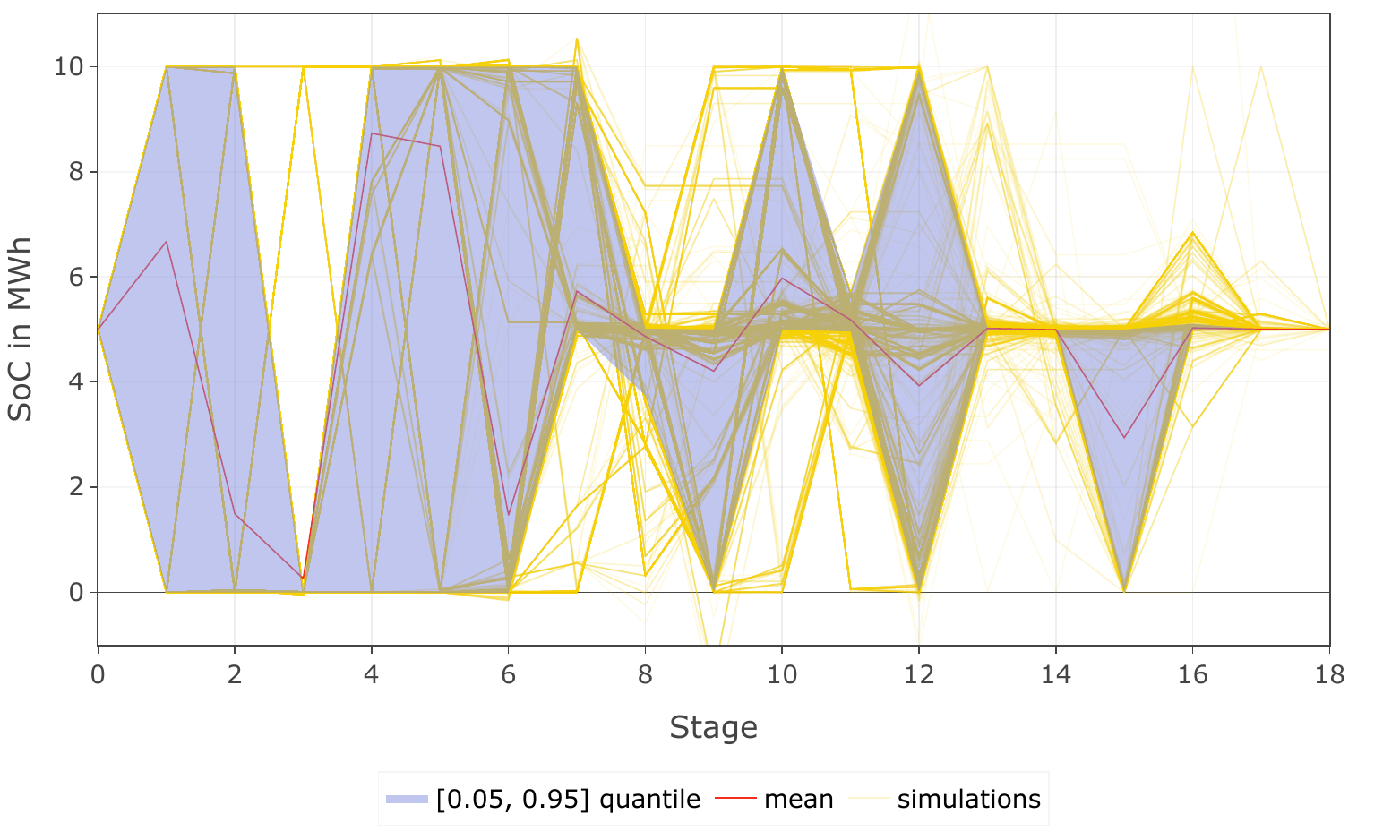}
        \caption{Battery SoC development at different stages.}
        \label{fig:SoC_over_stages_without_ID_const}
    \end{minipage}
\end{figure}

Convergence is reached with an upper bound at 10,500 \euro{} but with significant penalty terms, even at higher iteration counts. This leads to higher storage violations, in absolute occurrence and relative strength of the violation, as observable in Figure \ref{fig:iteration_plot_without_ID_const}. In Figure \ref{fig:SoC_over_stages_without_ID_const}, we can see more storage violations, especially small fluctuations up and down of the storage limits. We find the general trading patterns unchanged. 

\subsection{Price Paths of the Markov Chain in the Investigated Period}
\label{ch:scenarios_investigated_period}

\begin{figure}[htbp!]
\begin{minipage}{0.49\textwidth}
        \centering
        \includegraphics[width=\linewidth]{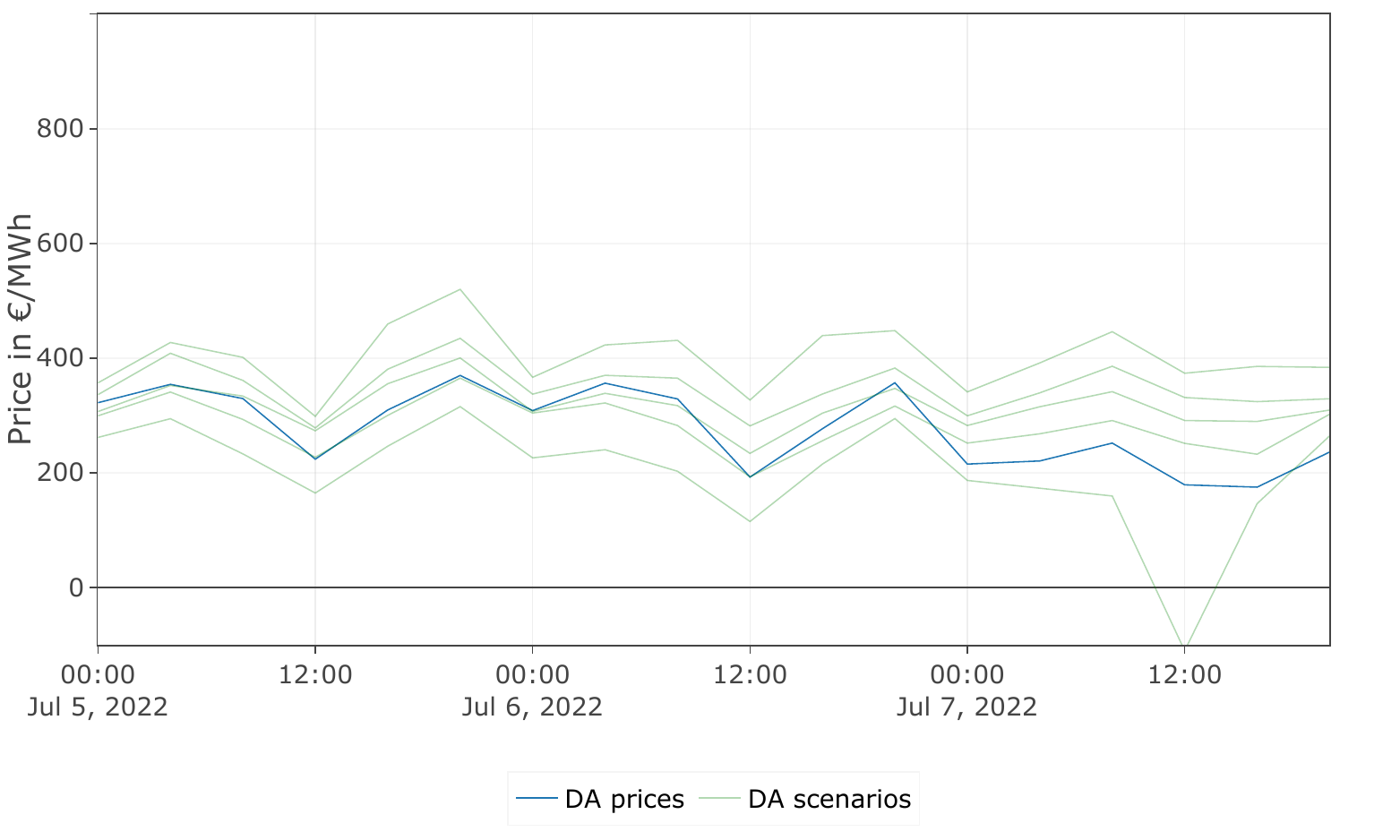}
        \caption{Price scenarios for the DA}
        \label{fig:scenarios_DA}
    \end{minipage}
    \hfill
    \begin{minipage}{0.49\textwidth}
        \centering
        \includegraphics[width=\textwidth]{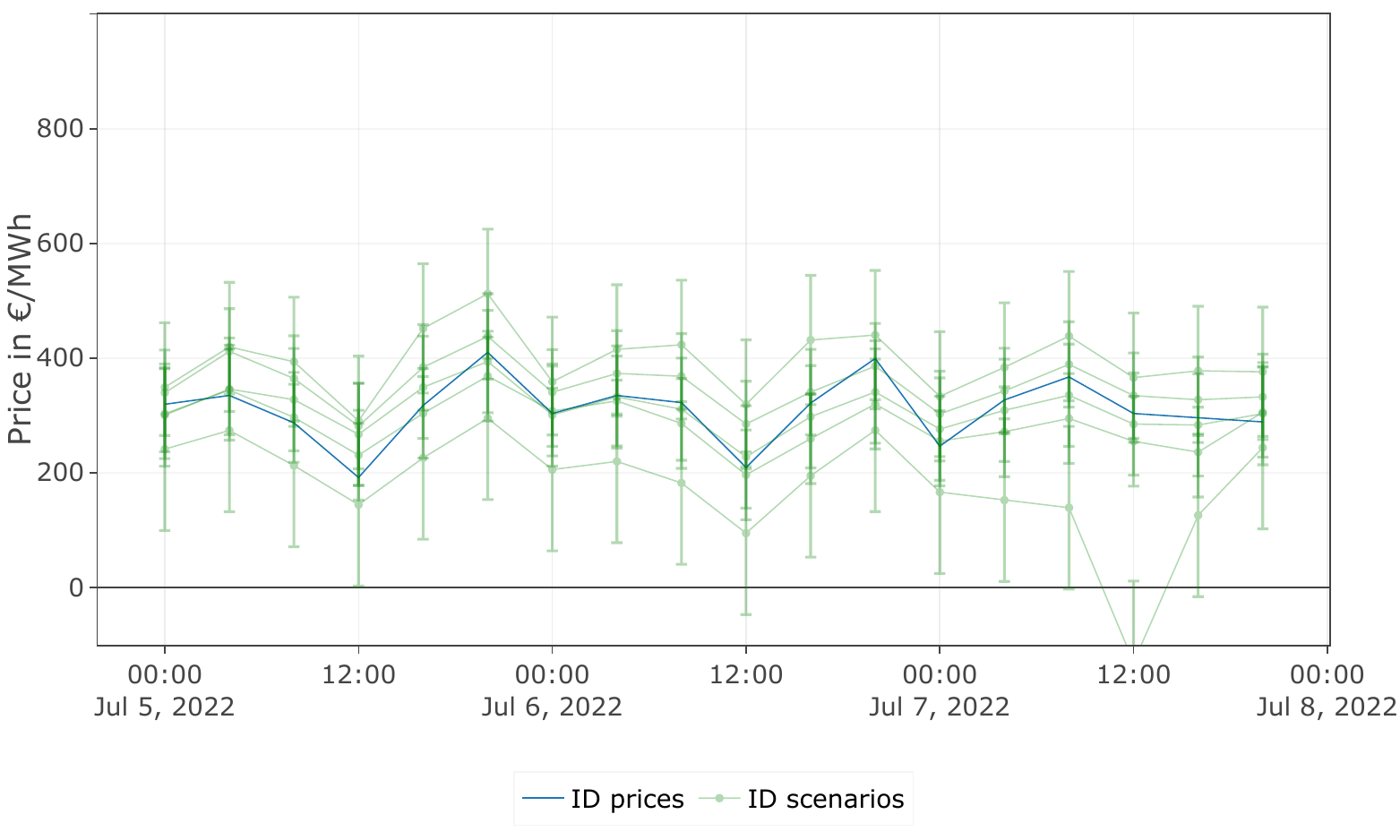}
        \caption{Price scenarios for the ID}
        \label{fig:scenarios_ID}
    \end{minipage}
\end{figure}
\begin{figure}[htbp!]
    \centering
    \includegraphics[width=0.5\linewidth]{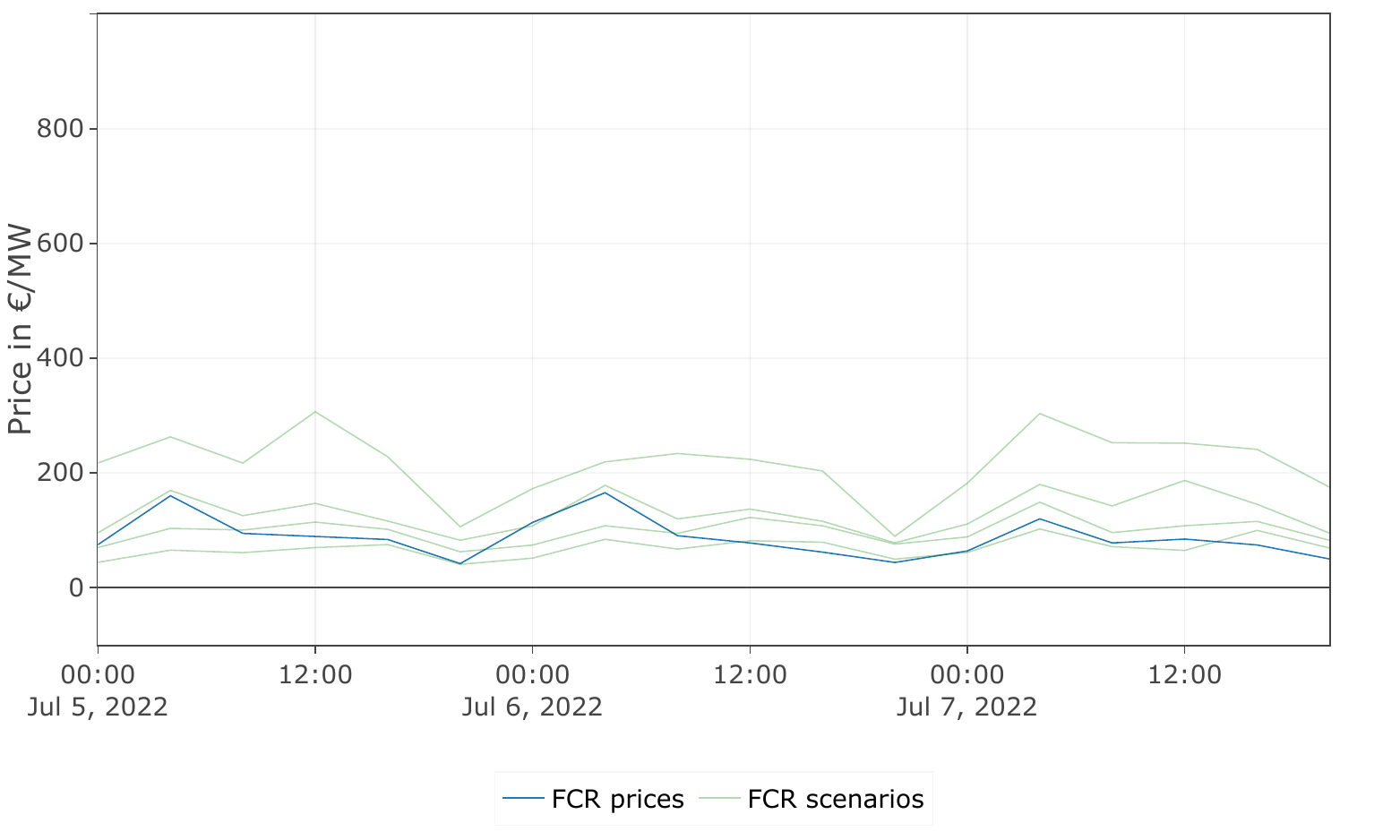}
    \caption{Price scenarios for the FCR}
    \label{fig:scenarios_FCR}
\end{figure}

\end{document}